\documentstyle[aps,multicol,eqsecnum,epsf]{revtex}
\begin{document}
\draft
\title{
Vortex statistics in a disordered two-dimensional XY model
}
\author{Lei-Han Tang}
\address{Institut f\"ur Theoretische Physik,
Universit\"at zu K\"oln, Z\"ulpicher Str. 77, D-50937 K\"oln, Germany\\
and\\
Condensed Matter Theory,
The Blackett Laboratory, Imperial College, London SW7 2BZ, United Kingdom
}
\date{\today}
\maketitle
\begin{abstract}
The equilibrium behavior of vortices in the classical
two-dimensional (2D) XY model with uncorrelated random phase shifts
is investigated.
The model describes Josephson-Junction
arrays with positional disorder, and has ramifications in
a number of other bond-disordered 2D systems.
The vortex Hamiltonian is that of a Coulomb gas
in a background of quenched random dipoles, which is
capable of forming either a dielectric insulator or a plasma.
We confirm a recent suggestion by
Nattermann, Scheidl, Korshunov, and Li
[J. Phys. I (France) {\bf 5}, 565 (1995)], and by
Cha and Fertig [Phys. Rev. Lett. {\bf 74}, 4867 (1995)]
that, when the variance $\sigma$ of random phase shifts is smaller
than a critical value $\sigma_c$, the system is in a phase 
with quasi-long-range order at low temperatures,
without a reentrance transition.
This conclusion is reached through a nearly exact calculation
of the single-vortex free energy, and a Kosterlitz-type
renormalization group analysis
of screening and random polarization effects from
vortex-antivortex pairs.
The critical strength of disorder $\sigma_c$ is found not to
be universal, but generally lies in the range $0<\sigma_c<\pi/8$.
Argument is presented to suggest that the system at $\sigma>\sigma_c$
does not possess long-range glassy order at any finite temperature.
In the ordered phase, vortex pairs undergo a series of 
spatial and angular localization processes as the temperature is lowered. 
This behavior, which is common to many glass-forming systems,
can be quantified through approximate
mappings to the random energy model and to the directed polymer on
the Cayley tree.
Various critical properties at the order-disorder transition are calculated.

\end{abstract}

\begin{multicols}{2}

\section{Introduction}

The Kosterlitz-Thouless-Berezinskii (KTB) transition
\cite{kost73,kost74,bere72} plays an important
role in the theory of ordering in two-dimensional (2D) systems
which has a continuous symmetry specified by a phase.
Examples include planar magnets, 2D solids, Josephson-Junction arrays,
and superfluid and superconductor films, etc.\cite{reviews}
These systems have an ordered phase at low temperatures, characterized
by power-law decay of correlations with distance.
The (quasi)long-range order is destroyed through unbinding of
vortex-antivortex pairs, which takes place at the KTB transition.

A question of both theoretical and practical interest 
is whether and how quenched disorder alters the above picture. 
In this paper we shall focus on the case of random frustration,
where disorder introduces random, uncorrelated phase shifts but do not 
pin the phase angles themselves. More precisely, we shall consider
an XY model with the following Hamiltonian\cite{rubi83},
\begin{equation}
H(\{\phi_i\})=-J\sum_{\langle ij\rangle}\cos(\phi_i-\phi_j-A_{ij}),
\label{hamil}
\end{equation}
where the sum runs over all nearest neighbor pairs on a square lattice.
The quenched random variables $A_{ij}$, which give a random bias
to the preferred advancing angle over each bond, are assumed to be
uncorrelated from bond to bond, and each is gaussian distributed
with the mean and variance given by
\begin{equation}
\langle A_{ij}\rangle=0,\quad
\langle A_{ij}^2\rangle=\sigma,
\label{A-corr}
\end{equation}
respectively.
It has been suggested that
model (\ref{hamil}) provides a good description of the 
Josephson-Junction arrays in a transverse magnetic 
field\cite{voss82,gran86,forr88,chak88,forr90}.
In this case, $\phi_i$ is identified with the phase
of the superconducting order parameter of grain $i$, and 
$A_{ij}=(2\pi/\Phi_0)\int_{i\rightarrow j}{\bf A}_{\rm ext}\cdot d{\bf l}$,
where ${\bf A}_{\rm ext}$ is the vector potential of the external
magnetic field and $\Phi_0=hc/2e$ is the superconducting flux quantum.
The case (\ref{A-corr}) corresponds to a situation where
the average magnetic flux over each elementary plaquette of
the grain network is an integer multiple of $\Phi_0$,
but random displacement of superconducting grains from a perfect
lattice structure yields quenched 
random phase shifts\cite{gran86,note1}.

On the theoretical side, model (\ref{hamil}) and its variants 
have been studied extensively in the past
\cite{rubi83,card82,nels83,pacz91,munt92,ozek93,kors93,nskl,cha95}.
Result of previous studies can be summarized as follows.
(i) The spin-wave fluctuations have the same excitation spectrum
as in the pure case. Disorder introduces distortion in the
ground state away from a perfect ferromagnetic alignment.
The combined effect of thermal and disorder fluctuations 
leads to an algebraic decay of the two-point phase-phase correlation function,
\begin{equation}
C_{\rm sw}(r_{ij})\equiv
\bigl\langle\exp\bigl[i(\phi_{{\rm sw},i}-\phi_{{\rm sw},j})\bigr]\bigr\rangle
\sim r_{ij}^{-\eta_{\rm sw}},
\label{2pt}
\end{equation}
where $r_{ij}$ is the distance between site $i$ and $j$,
and 
\begin{equation}
\eta_{\rm sw}={1\over 2\pi}\bigl({T\over J}+\sigma\bigr)
\label{eta}
\end{equation}
is the correlation length exponent at temperature $T$,
due to spin-waves only.
(ii) Vortices, which are topological point defects in the $\phi$-field,
interact with each other and with the quenched disorder
through a Coulomb potential.
The interaction between two vortices is of the charge-charge type,
where the charge of each vortex is given by its vorticity.
The interaction between a vortex and a particular
disordered bond is of the charge-dipole type, with the strength
of the dipole given by the phase shift $A_{ij}$ over the bond.
The equilibrium statistics of vortices is decoupled from that of spin waves.
For a long time, the phase diagram of the model was thought to be of the
type illustrated in Fig. 1(a)\cite{rubi83,card82}.
For $\sigma<\sigma_c\simeq \pi/8$, a phase with bound vortex-antivortex
pairs, and hence algebraic decay of phase correlations,
still exists, but only in a temperature window
$T_-(\sigma)<T<T_+(\sigma)$.
Below $T_-(\sigma)$, a ``re-entrant'' disordered phase was predicted.
The two transition temperatures coincide at a critical
strength of the disorder $\sigma_c$, above which
the ordered phase disappears altogether.
Two recent papers, by Nattermann, Scheidl, Korshunov and Li (NSKL)\cite{nskl},
and by Cha and Fertig\cite{cha95}, cast doubt on the reentrance picture.
The phase diagram they suggested is shown in Fig. 1(b),
where the reentrance line $T_-(\sigma)$ disappears.
NSKL\cite{nskl,kors95,sche96} further suggested 
that some sort of freezing phenomenon
takes place below a certain temperature 
\begin{equation}
T_\ast(\sigma)=2\sigma J
\label{T-star1}
\end{equation}
[see the dashed line in Fig. 1(b)],
which preempts the reentrance transition at $T_-(\sigma)<T_\ast(\sigma)$ 
found previously.

The aim of the present paper is to expand the pioneering ideas presented
in Refs. \cite{nskl} and \cite{cha95}
to unfold the physics which underlies the vortex-antivortex
unbinding transition in the presence of the quenched disorder.
There are two main extensions contained in this work as detailed below.

First, we analyze quantitatively the equilibrium behavior of
a single vortex in a background of quenched random dipoles.
Analogy is made to two well-studied problems
involving disorder: the random energy model\cite{derr81}, and a
directed polymer on the Cayley tree\cite{derr88}.
It is shown that the single-vortex problem has a 
glass transition at a temperature 
\begin{equation}
T_g=J(\pi\sigma/2)^{1/2},
\label{Tg}
\end{equation}
below which entropy goes to zero, i.e., the vortex becomes localized
at the lowest energy site. The free energy of the vortex is proportional
to the logarithm of system size at all temperatures,
with a prefactor which vanishes on the phase boundary shown in Fig. 1(b).

Second, the dielectric and freezing properties of
a dilute gas of vortex-antivortex pairs (or molecules)
are examined in further detail,
with particular emphasis on the spatial structure of
equilibrium pair configurations.
The freezing line $T=T_\ast$ in Fig. 1(b)
is shown to be related to the loss of entropy of a pair
over an area where the pair can be considered as isolated
from other pairs of comparable size.
If we fix the center position of the pair,
the two vortices make up the pair freeze at $T_g$.
In the ordered phase, $T_\ast<T_g$ due to the fact
that the pair is allowed to explore an area much larger
than its size and hence has a lower freezing temperature.
Interestingly, freezing of pairs is not associated
with a singularity in the free energy of the system as a whole,
and there is no real phase transition at $T_\ast$.
Disorder also generates random, zero-field polarization of 
the gas of pairs, which enhances the effective disorder seen by vortices
separated by a large distance.
This effect, which has been previously overlooked,
shifts the horizontal phase boundary $\sigma=\pi/8$ in Fig. 1(b)
to smaller values of $\sigma$\cite{sche96}. 

The outcome of these considerations can be turned into
a set of renormalization group (RG) recursion relations
which capture the {\it average, large-distance} properties of the system.
Apart from some minor differences, the RG flow equations derived in
this paper are in agreement with those of Ref.\cite{nskl}.
To the extent that such a simplifying description
offers a good approximation, a phase diagram of the
form Fig. 1(b) is produced.
The RG description is however not sensitive enough to rare fluctuations.
The influence of rare fluctuations on some of the quantitative aspects
of our results, such as the slope of the phase boundary
as $T$ tends to zero, remain to be studied.
Qualitatively, though, the basic conclusions of the
RG calculation are expected to be valid, as the modification
of the bare interactions due to excitation of large-size pairs is 
relatively small in the entire ordered phase shown in Fig. 1(b).

An interesting question is whether the system in the low temperature
region above the $\sigma_c$-line has long-range glassy order.
Our calculation of the dielectric susceptibility of a gas of
pairs indicates that screening is present at all temperatures,
despite localization in the orientation of individual pairs
below $T_g$. This supports the idea that, in the disordered phase,
vortex-vortex interaction at large distances is
always short-ranged. Consequently, long-range glassy order in the 
phase field is not expected at any nonzero temperature
due to finite energy cost to excite an additional vortex in the system.

The paper is organized as follows.
In Sec.~II the Coulomb gas representation of vortices of the XY model
is briefly reviewed. A qualitative discussion of 
vortex-antivortex unbinding is presented to highlight the 
outstanding issues.
The problem of a single vortex interacting with quenched random
dipoles is analyzed in Sec.~III. Connection is made to the
random energy model and to a directed polymer on the Cayley
tree. In Sec.~IV we examine the behavior of a dilute gas of
vortex-antivortex pairs of comparable size, under the influence of disorder.
The calculation of the dielectric susceptibility
and the zero-field polarization of such a gas is presented, 
as well as an analysis of fluctuations of pair density.
A physical interpretation of the $T_\ast$ line is proposed.
Section V contains a derivation of the RG recursion relations
and results that follow from these equations.
A discussion of the phase diagram, singularity of the free energy,
divergence of the correlation length, and the two-point phase-phase
correlation function is presented.
The main results of the paper are summarized in Sec.~VI.
Some of the technical aspects of the study are relegated to
the four appendices at the end.

\section{Coulomb gas formulation}

\subsection{Vortex Hamiltonian}

To set the stage, let us review briefly the steps leading
to the Coulomb gas representation of (\ref{hamil}).
The standard procedure is to take a
continuum limit of (\ref{hamil}), which yields
a quadratic Hamiltonian\cite{vill75,rubi83},
\begin{equation}
H={J\over 2}\int d^2r[\nabla\phi-a^{-1}{\bf A}({\bf r})]^2,
\label{quadr-h}
\end{equation}
where $a$ is the lattice constant. The two components of
{\bf A} are given by the disorder $A_{ij}$ on adjacent
horizontal and vertical bonds, respectively.

In the presence of vortices, the field $\phi({\bf r})$ is multi-valued.
The vortex configuration is specified by a set of 
vortex charges $m_i$ such that the phase advance along a closed
path surrounding site $i$ (or rather cell $i$) is given by
$$\oint d\phi =2\pi m_i.$$
In a system with periodic boundary conditions,
neutrality $\sum_i m_i=0$ is satisfied.
The gradient of the $\phi$ field can now be decomposed into
a rotation-free part and a divergence-free part,
\begin{equation}
\nabla\phi=\nabla\phi_{\rm sw}+\sum_i m_i\hat{\bf z}\times({\bf r}-{\bf r}_i)
/|{\bf r}-{\bf r}_i|^2,
\label{phi_dec}
\end{equation}
where $\phi_{\rm sw}$ represents ``spin-wave'' fluctuations.
The same procedure can be repeated for {\bf A},
\begin{equation}
{\bf A}=a\nabla\phi_0+{\bf A}_{\rm r},
\label{A_dec}
\end{equation}
where the potential $\phi_0$ satisfies 
\begin{equation}
a\nabla^2\phi_0=\nabla\cdot{\bf A}.
\label{phi0}
\end{equation}

Inserting Eqs. (\ref{phi_dec}) and (\ref{A_dec}) into (\ref{quadr-h}),
we obtain (apart from a constant) 
$H=H_{\rm sw}+H_{\rm v}$, where the spin-wave part is given by
\begin{equation}
H_{\rm sw}={J\over 2}\int d^2r(\nabla 
\phi_{\rm sw}-\nabla\phi_0)^2,
\label{spinwave}
\end{equation}
and the vortex part given by
\begin{equation}
H_{\rm v}=\sum_i (m_i^2 E_c+ m_iV_i)
-\pi J\sum_{i\neq j}m_im_j\ln{r_{ij}\over a}.
\label{c-gas}
\end{equation}
(See Appendix~A for more details on the derivation.)
Here and elsewhere ${\bf r}_{ij}={\bf r}_i-{\bf r}_j$ is the
displacement vector between sites $i$ and $j$,
and $r_{ij}=|{\bf r}_{ij}|$ is the distance. 
In addition to the usual core energy $E_c$,
a vortex interacts with a quenched random dipole field
${\bf q}_i=(a/2\pi){\bf A}({\bf r}_i)\times\hat{\bf z}$
through the potential
\begin{equation}
V_i\equiv V({\bf r}_i)= 
2\pi J\sum_{j\neq i}{\bf q}_{\; j}\cdot{\bf r}_{ij}/r_{ij}^2.
\label{dipole}
\end{equation}
From the above definition we have, in component form,
\begin{equation}
\langle q_{i,\alpha}\rangle=0,\qquad
\langle q_{i,\alpha}q_{j,\beta}\rangle=(a/2\pi)^2\sigma
\delta_{ij}\delta_{\alpha\beta}.
\label{pp-cor}
\end{equation}
Note that $V_i$ vanishes when {\bf A} is rotation-free.

\subsection{Pair-unbinding transition}

At sufficiently high core energies, at least, the
gas of vortices in a charge-neutral system is
expected to form one of the two phases described below.
The first is a dielectric insulator, where
$\pm 1$ charges bind to form pairs of charge-neutral molecules.
This structure is low in the Coulomb energy, but also low in
entropy due to binding.
The second is a plasma with a finite density of unpaired 
(or free) vortices. This structure is high in the Coulomb energy
but also high in entropy.
In the absence of disorder, both the Coulomb energy and entropy
scale logarithmically with distance in two dimensions.
A simple energy-entropy argument then predicts a
finite temperature transition for the unbinding of 
vortex-antivortex pairs. This is also the temperature where
the free energy of a single vortex goes to zero.
An improved treatment, which takes into account 
reduction of the Coulomb energy due to screening by
other vortex-antivortex pairs, yields an exact description
of the critical properties at the transition.
In the plasma phase, there is complete screening of
the Coulomb potential, so that interaction between distant
charges become short-ranged.

In the presence of quenched random dipoles, vortices may explore fluctuations
in the disorder potential to lower their Coulomb energy, and hence
become more numerous. This speaks for the reduced stability of the
insulating phase. On the other hand, in the process of gaining potential
energy, vortices become more localized, and this way loose entropy.
The first insight one needs is how much energy a vortex can gain
from the disorder by positioning itself at the right place.
It turns out that this problem can be solved almost exactly,
and the result again has logarithmic scaling with distance.
The amplitude of energy gain from disorder is proportional to $\sigma^{1/2}$
at low temperatures.
Thus, when entropy is not a factor, excitation of free vortices 
is not expected below a certain critical strength of the disorder.

As in the pure case, a complete treatment requires analysis of
screening of the Coulomb potential due to other pairs of vortices
present in the system. 
At high temperatures, a pair is able to explore a large
number of different disorder environment, which minimizes
the difference between quenched and annealed disorder.
The situation becomes different at low temperatures where,
as in the random energy model, the equilibrium behavior
of a pair is dominated by the lowest energy configuration
in the area accessible to the pair.
A crucial issue is thus to obtain the correct statistics of
the pair when spatial and angular localization becomes important.

With the above general picture in mind, we are in a position to
perform the necessary calculations.

\section{Single vortex}

In this section, we examine the behavior of a single vortex, confined in
a box of linear dimension $R\gg a$.
In the presence of disorder, the energy of the vortex depends
on its position $i$,
\begin{equation}
E_i=E_c+\pi J\ln(R/a)+V_i,
\label{s-eng}
\end{equation}
where $V_i$ is given by (\ref{dipole}) with the sum restricted
to sites in the box. 
The variance and spatial correlations of $V_i$ are given by
\begin{eqnarray}
\langle V_i^2\rangle&=&2\pi\sigma J^2\ln(R/a)+O(1),\\
\label{v_i}
\\
\langle(V_i-V_j)^2\rangle&=&4\pi \sigma J^2\ln(r_{ij}/a)+O(1).
\label{vv}
\end{eqnarray}

A simplifying approximation to the single-vortex problem is obtained
by setting the correlation of $V_i$ to zero.
The resulting problem is known as the {\it random energy model} 
(REM)\cite{derr81}.
It turns out that, for quantities of interest to us, 
correlations in the disorder potential only introduce 
minor corrections to the REM results.
In the following we shall first discuss the REM 
and then an improved representation.

\subsection{Random energy approximation}

In the REM one considers the partition function
\begin{equation}
z=\sum_{i=1}^N\exp(-x_i/T),
\label{rem-partition}
\end{equation}
where $x_i, i=1,\ldots, N$, are a set of random energy levels drawn
independently from a gaussian distribution,
\begin{equation}
\psi(x)=(2\pi \Delta)^{-1/2}\exp(-x^2/2\Delta).
\label{rem}
\end{equation}
The model has been analyzed in great detail by Derrida\cite{derr81}.
Below we quote some of his results relevant for our discussion,
and refer the reader to his original paper for further 
details. (See also Appendix~B.)

In the thermodynamic limit $N\rightarrow\infty$ while fixing
the ratio $s\equiv \Delta/\ln N$,
the average free energy is extensive in $\ln N$,
\begin{equation}
\langle f\rangle\equiv -T\langle\ln z\rangle=-c(T,s)\ln N+O(\ln\ln N),
\label{rem-f}
\end{equation}
where 
\begin{equation}
c(T,s)=
\left\{
\begin{array}{l}
T+s/(2T),\;\;\mbox{for}\; T>T_g(s);\\
\\
(2s)^{1/2},\;\;\mbox{for}\; T<T_g(s).
\end{array}\right.
\label{c-rem}
\end{equation} 
Here 
\begin{equation}
T_g(s)=(s/2)^{1/2}
\label{T_g-rem}
\end{equation}
is the freezing temperature of the model.
For $T<T_g$, the entropy is no longer extensive in $\ln N$.

The above result can be applied to the single vortex problem by
substituting $N\rightarrow (R/a)^2$, 
$\Delta\rightarrow 2\pi\sigma J^2\ln(R/a)$,
and $s\rightarrow \pi\sigma J^2$.
From (\ref{rem-f}), we obtain the average free energy of the vortex,
\begin{equation}
\langle F\rangle\simeq 
\left\{
\begin{array}{l}
\displaystyle{
E_c+\pi J\bigl(1-{2T\over \pi J}-{\sigma J\over T}\bigr)\ln{R\over a},
}
\;\;\mbox{for}\; T>T_g;\\
\\
\displaystyle{
E_c+\pi J\bigl(1-\sqrt{{8\sigma\over\pi}}\bigr)\ln {R\over a},
}
\;\;\mbox{for}\; T<T_g.
\end{array}\right.
\label{f_sing}
\end{equation}
The corresponding freezing temperature is given by
Eq. (\ref{Tg}) (solid line in Fig. 2).
The coefficient of the logarithm changes sign across the 
dashed line shown in Fig. 2, which is precisely the phase
boundary in Fig. 1(b) when 
renormalized values for $J$ and $\sigma$ are used.
(See discussion in Sec.~V.)

Below $T_g$, the entropy of the vortex is no longer extensive
in $\ln(R/a)$. In fact, it can be shown that 
only one or a few lowest energy sites contribute significantly
to the partition sum (\ref{rem-partition}) in this regime
(see Appendix~B).
Within the region bounded by the dashed line in Fig. 2,
the typical free energy of a vortex is positive, but
there are rare realizations of disorder which
give rise to a negative free energy.
The probability for such events is a power-law function of
$R/a$ with a negative exponent.
This fact is important when we consider pair excitations in
Sec.~IV.

\subsection{Correlations in the disorder potential}

The REM approach to the single-vortex problem is not completely
satisfactory as it ignores spatial correlations in the energies $V_i$.
This correlation has a simple origin (see Fig. 3). 
When we move the vortex from a site $i$
to a site $j$, the change in the disorder potential is mainly
due to a change in the local environment up to a distance of
order $r_{ij}$, as contributions to $V_i$ and $V_j$
from quenched dipoles further away are nearly identical.
This type of correlation can be easily coded using the Cayley tree,
where each site is associated with a path on the tree.
The potential on a site is made equal to the energy of a path on the
tree. Geometrical proximity is translated into hierarchical proximity
on the tree.

This representation can be made explicit using the following construction,
though details of it should be unimportant for our conclusions.
For any chosen site $i$, we divide the space into a set of rings
of inner and outer radii $R_{n-1}$ and $R_n$, respectively,
such that $a=R_0<R_1<\ldots<R_m\simeq R$, while keeping
$R_n/R_{n-1}=b$ constant.
The potential at site $i$ can be written
as a one-dimensional sum, $V_i=\sum_n V_i^{(n)}$, 
where each term in the sum contains only contributions from dipoles within
a given ring, i.e.,
\begin{equation}
V_i^{(n)}=2\pi J
\sum_{R_{n-1}\leq r_{ik}<R_n}{\bf q}_k\cdot {\bf r}_{ik}/r^2_{ik}.
\label{V-n}
\end{equation}
We now identify the $n$th ring with the $n$th node (branching point)
along the path $i$ on the tree, where $n$ increases from bottom to top.
The energy of the node is given by $V_i^{(n)}$.
Repeating the above procedure for a different site $j$,
we obtain another sequence of energies $V_j^{(n)}$
for nodes on the path $j$. The two paths join on level 
$n_{ij}=\ln(r_{ij}/a)/\ln b$.

An intriguing fact about the random dipolar interaction is that
the subsums constructed above are gaussian random variables 
with identical statistics,
\begin{equation}
\langle V_i^{(n)}\rangle=0,\quad
\langle V_i^{(n)}V_i^{(n')}\rangle=2\pi\sigma J^2(\ln b)\delta_{n,n'}.
\label{subsum}
\end{equation}
Thus all rings contribute equally to the sum $V_i$,
independent of the radius of the ring.

The Cayley tree problem discussed above has been analyzed
in detail by Derrida and Spohn\cite{derr88}. 
Its properties are quite similar to the REM.
In particular, the extensive part of the free energy 
is the same as in the REM, independent of the choice of $b$.
In addition, moments of the partition function
have the same dependence on $N$ as indicated in Eq. (\ref{z-n}),
and the transition temperature $T_n$ of the $n$-th moment
is the same as in the REM.
There are, however, differences in the amplitude of the
ratio $\langle z^n\rangle/\langle z\rangle^n$.
This implies that the distribution of the free energy,
$f=-T\ln z$, is not exactly given by Eqs. (\ref{rho-min}) and
(\ref{rho-minapp}) for $f$ significantly less than $\langle f\rangle$,
but the difference should be small, as otherwise 
the behavior of $\langle z^n\rangle$ would be significantly different.

\section{Dilute gas of paired vortices}

As mentioned in Sec.~II.~B,
a quantitative study of the pair-unbinding transition
must include a discussion of pair-excitations which
modify the Coulomb interaction at large distances.
This is usually done by employing a real-space RG
procedure, to be explained in detail in Sec.~V. 
A crucial step in the RG scheme is the calculation
of the dielectric susceptibility and zero-field polarization
of a gas of pairs in a certain size range, say between 
$R$ and $R+dR$. This is the task to be carried out in this section.

\subsection{Lattice gas representation}

To treat a dilute gas of pairs of uniform size $R$,
it is useful to separate the ``internal'' degrees of freedom
of a pair, given by allowed configurations of the pair confined
to a box of linear size $R$, from rigid translations of the pair
over a distance greater than $R$.
One way of implementing the idea is to 
impose a lattice with a lattice constant $R$.
The lattice-gas representation is extremely handy
owing to the following two properties of the system:
(i) the disorder potential on a pair is 
essentially uncorrelated when the pair is translated 
over a distance larger than $R$;
(ii) interaction between pairs of similar 
size in the dilute limit can be approximated by
a hard-core potential extending
over a distance of the pair size $R$.
These facts can be established following a similar line
of reasoning as in the original paper by Kosterlitz
and Thouless\cite{kost73}.

Let ${\bf r}^+$ and ${\bf r}^-$ be the coordinates of $+1$ and
$-1$ charges in a pair, respectively.
The pair energy is given by
\begin{equation}
E_p=2E_c+2\pi J\ln(R/a)+V({\bf r}^+)-V({\bf r}^-),
\label{E-pair}
\end{equation}
where $R=|{\bf r}^+-{\bf r}^-|$ is the size of the pair.

The rapid decay of correlations in the disorder potential
$V({\bf r}^+)-V({\bf r}^-)$ on a pair beyond a distance of order $R$
comes from an observation made in Sec.~III.~B.
The two charges which make up a pair interact separately with
quenched random dipoles within a distance of order $R$ from the
pair center, but collectively as a dipole when more distant disorder
is in question.
Hence the random part of $E_p$ is dominated by disorder
within a distance of order $R$ from the pair center.
(The remaining contribution from distant quenched random dipoles
can be treated as a perturbation when necessary.)
On the other hand, barring contributions from distant quenched dipoles,
$V({\bf r}^+)$ is quite independent from $-V({\bf r}^-)$
for two reasons. First, each potential is dominated by
quenched dipoles in the immediate vicinity of the site in question
(see discussion on the ring structure in Sec.~III.~B).
Second, although the two charges are in the same disorder environment,
when it comes to optimizing their (free) energies, they see opposite
ends of the disorder energy distribution due to the difference in sign.
Therefore, to a good approximation, we can replace $E_p$ by
the sum of two single vortex energies of the form (\ref{s-eng}),
each containing a random potential generated by quenched dipoles
within a box of linear size $R$, independent from the other.

The interaction between one pair and another is of the
dipole-dipole form at large distances, which is
small compared to $E_p$ and can be treated as a perturbation.
The interaction becomes more complex when two pairs are at 
a distance $R_1<R$, but it is generally repulsive, 
with a strength of order $4\pi J\ln(R/R_1)$.
(Note that the two pairs should be arranged in such a way that
it is not possible to regroup them to form $\pm 1$ pairs of smaller sizes.)
For simplicity, we shall
replace the interaction by a hard-core potential of range $R$. 
In the dilute limit, the main effect of this interaction is
to prevent more than one pair to take advantage of a particular
favorable configuration (and the ones very close to it),
which turns out to be a very important constraint 
at low temperatures\cite{nskl}.

We are now in a position to define the lattice-gas representation.
We divide the plane into a square lattice of cells,
each of linear dimension $R$.
Any given cell has at most one pair, and pairs in different
cells do not interact with each other.
The Boltzmann weight on an occupied cell
can be written as $y_pz_p$, where 
\begin{equation}
y_p\equiv (R/a)^{-2\pi J/T}\exp(-2E_c/T)
\label{yp}
\end{equation}
is the {\it pair fugacity} and
$z_p$ is the configurational partition function
of the pair attached to the cell.
Since there is no interaction between different cells,
the partition function of the system factorizes into
a product of cell partition functions $1+y_pz_p$.
In addition, average over all cells can be replaced by
an average over the disorder, as each cell represents
an independent realization.

To apply the lattice-gas description to
the system of pairs in a given size range, say between $R$ and $R+dR$,
we need to specify $z_p$ in more detail.
For the discussion to be meaningful, $dR$ should be small enough
so that the pair fugacity $y_p$ can be regarded as a constant, but
large enough so that individual charges in a pair are allowed
to explore their own local disorder environment without been
severely constrained by the specified range of pair size.
Both criteria can be met by choosing $dR\sim R$.
The configurational partition function of an occupied cell is given by
\begin{equation}
z_p=\sum_{\scriptstyle({\bf r}^++{\bf r}^-)/2\in {\rm cell}\atop
\scriptstyle R\leq |{\bf r}^+-{\bf r}^-|<R+dR}
\exp\bigl[-{V({\bf r}^+)-V({\bf r}^-)\over T}\bigr].
\label{z-p}
\end{equation}
The potential $V({\bf r}^+)-V({\bf r}^-)$ inside a cell
has a spatial correlation
of similar nature as the potential on a single vortex discussed in
Sec.~III. To simplify the calculation, we shall again make
the random energy approximation where this correlation is ignored.
The parameters of the REM applied to the 
problem of pairs are, 
\begin{equation}
N=2\pi(R/a)^4(dR/R),\quad\Delta=4\pi\sigma J^2\ln (R/a).
\label{rem-pair}
\end{equation}
For $dR\simeq R$, the freezing temperature $T_g$ for the pair
in a cell is the same as the freezing temperature of a single
vortex, Eq. (\ref{Tg}).

\subsection{Pair density}

In equilibrium, the probability of finding a pair in a given cell
is given by
\begin{equation}
W={y_pz_p\over 1+y_pz_p}.
\label{W}
\end{equation}
For a dilute gas, the typical value of $W$
is given by $W_{\rm typ}\simeq y_pz_{p,\rm typ}$,
where $z_{p,\rm typ}$ is the typical value of $z_p$
(see discussion in Appendix~B).
Combining Eqs. (\ref{rem-f}), (\ref{c-rem}), (\ref{z-typ1}), and (\ref{yp}),
we obtain,
\begin{equation}
W_{\rm typ}\sim
\left\{
\begin{array}{l}
\displaystyle{
\Bigl({R\over a}\Bigr)^{4-2\pi K+2\pi\sigma K^2}{dR\over R},
}
\;\;\mbox{for}\; T>T_g;\\
\\
\displaystyle{
\Bigl({R\over a}\Bigr)^{-2\pi K(1-\sqrt{8\sigma/\pi})}{dR\over R},
}
\;\;\mbox{for}\; T<T_g.
\end{array} \right.
\label{W-typ}
\end{equation}
Here $K\equiv J/T$.
The exponent of the power-law changes sign on the
dashed line in Fig. 2.

Like $z_p$, $W$ has a broad distribution. 
Its mean value $\langle W\rangle$
deviates significantly from $W_{\rm typ}$ for $T<T_g$. 
Since the $n$-th moment of $z_p$ grows much faster than 
$\langle z_p\rangle^n$ for sufficiently large $n$, 
it is not possible to calculate $\langle W\rangle$
by expanding the right-hand-side of (\ref{W}) as a power series of $y_pz_p$.
Nevertheless, the average can be calculated by
treating the cases $y_pz_p<1$ and $y_pz_p>1$ separately,
as done in Appendix~C. Results of the calculation
are given by Eqs. (\ref{W-mean1a}) and (\ref{W-mean2a}) in respective
temperature regimes.
For $R\gg a$, a power-law dependence of $\langle W\rangle$ on $R$
is found, 
\begin{equation}
\langle W\rangle\sim
\left\{
\begin{array}{l}
\displaystyle{
\Bigl({R\over a}\Bigr)^{4-2\pi K+2\pi\sigma K^2}{dR\over R},
}
\;\;\mbox{for}\; T>T_\ast;\\
\\
\displaystyle{
\Bigl({R\over a}\Bigr)^{4-\pi/(2\sigma)}{dR\over R},
}
\;\;\mbox{for}\; T<T_\ast.
\end{array} \right.
\label{W-mean_a}
\end{equation}
The exponent freezes to a temperature-independent
value below $T_\ast$.

\subsection{Zero-field polarization}

The disorder environment in a given cell specifies a
favorable configuration for a pair in the cell.
The breaking of rotational invariance thus yields a
zero-field dipole moment,
\begin{equation}
{\bf p}_0\equiv 
{\sum {\bf p}\exp[-E_p/T]
\over 1+y_pz_p}
\label{p0}
\end{equation}
where ${\bf p}={\bf r}^+-{\bf r}^-$ is the dipole moment of the pair.
The sum in Eq. (\ref{p0}) is restricted to the internal degrees
of freedom of the pair, as in Eq. (\ref{z-p}).

Due to statistical rotational symmetry,
$\langle{\bf p}_0\rangle=0$. 
Its variance can be calculated approximately from the following
consideration.
Note that ${\bf p}_0$ is small when many distinct
configurations contribute to the cell partition sum $z_p$.
It becomes large when the lowest energy configuration
(and nearby configurations with approximately the same
orientation of {\bf p}) dominates.
Based on the discussion of Appendix~B, it is reasonable to assume
that the latter occurs whenever $z_p$ is significantly
larger than its typical value, $z_{p,\rm typ}$. 
Replacing {\bf p} inside the sum in (\ref{p0})
by the dipole moment of the ground state,
we make an error with a probability of the order of
$W_{\rm typ}$, which is smaller than $\langle W\rangle$.
This yields the estimate,
\begin{equation}
\langle|{\bf p}_0|^2\rangle/R^2=\langle W^2\rangle+
O\bigl(\langle W\rangle^2\bigr).
\label{p0-var}
\end{equation}
The calculation presented 
at the end of Appendix~C yields, for $T<T_\ast$,
\begin{equation}
\langle W^2\rangle\simeq (1-T/T_\ast)\langle W\rangle.
\label{p0-1}
\end{equation}
For $T>T_\ast$, $\langle W^2\rangle$ decays faster with $R$
than $\langle W\rangle$.
The distribution of $|{\bf p}_0|$ is expected to be broad.
In particular, for $T<T_g$, where typically one or two
configurations dominate the partition sum,
the distribution of $|{\bf p}_0|$ is similar to the distribution of $W$.

Let us now consider the
correlation between ${\bf p}_0$ and the total dipole moment of
disorder in the cell,
\begin{equation}
{\bf q}=\sum_{i\in\rm cell}{\bf q}_i.
\label{q-cell}
\end{equation}
Since ${\bf p}_0$ is mostly determined by the arrangement of
the disorder in the immediate vicinity of the two charges
making up the pair, we expect the contribution
to ${\bf p}_0$ from ${\bf q}$ to be small,
but the effect is important for later discussions.
To estimate the contribution,
let us consider a quantity $\tilde{\bf p}_0$,
which is the equivalent of ${\bf p}_0$
under the replacement
${\bf q}_i\rightarrow \tilde{\bf q}_i={\bf q}_i-(a/R)^2{\bf q}$.
From the third example of Appendix~A, we see that switching on
${\bf q}$ is equivalent to switching on a polarizing field
${\bf E}_q=-2\pi^2 J{\bf q}/R^2$.
Linear response theory then suggests, on average,
a relation of the form,
\begin{equation}
{\bf p}_0\simeq \tilde{\bf p}_0-2\pi^2\bar\chi J{\bf q},
\label{p0-split}
\end{equation}
where $\bar\chi$ is the average dielectric susceptibility
of the gas of pairs, to be discussed below.

\subsection{Induced polarization}

In the presence of a weak, constant external electric field {\bf E},
a cell acquires an induced dipole moment due to pair excitation,
\begin{equation}
{\bf p}_{\rm ind}=
{\sum {\bf p}\exp[-(E_p-{\bf p}\cdot{\bf E})/T]
\over 1+\sum\exp[-(E_p-{\bf p}\cdot{\bf E})/T]}-{\bf p}_0.
\label{p-ind}
\end{equation}
The induced polarization ${\bf P}_{\rm ind}$ of the gas of pairs
is given by the spatial average of ${\bf p}_{\rm ind}$, or equivalently,
the disorder average,
\begin{equation}
{\bf P}_{\rm ind}=R^{-2}\langle {\bf p}_{\rm ind}\rangle.
\label{P-ind}
\end{equation}
To the first order in {\bf E}, 
we find 
\begin{equation}
{\bf P}_{\rm ind}=\bar\chi {\bf E}, 
\label{chi}
\end{equation}
with
\begin{equation}
\bar\chi=(2T)^{-1}\bigl(\langle W\rangle
-\langle|{\bf p}_0|^2\rangle/R^2\bigr).
\label{chibar}
\end{equation}
Using Eqs. (\ref{W}) and (\ref{p0-var})
we may rewrite the above equation as,
\begin{equation}
\bar\chi={1\over 2T}\Bigl[y_p{\partial\langle W\rangle\over\partial y_p}
+O\bigl(\langle W\rangle^2\bigr)\Bigr].
\label{chibar1}
\end{equation}
Here we have used the identity
\begin{equation}
y_p\partial \langle W\rangle/\partial y_p=\langle W\rangle-\langle W^2\rangle.
\label{partial-W1}
\end{equation}

The derivative in the above equation can be evaluated
using Eq. (\ref{W-mean1}) for $T>T_\ast$,
and (\ref{W-mean2}) for $T<T_\ast$.
To leading order, the result reads,
\begin{equation}
\bar\chi\simeq
\left\{
\begin{array}{l}
(2T)^{-1}\langle W\rangle,\;\;\;\mbox{for}\; T>T_\ast;\\
\\
(2T_\ast)^{-1}\langle W\rangle,
\;\;\mbox{for}\; T<T_\ast.
\end{array}\right.
\label{chibar2} 
\end{equation}
[Note that, in both cases, the coefficient in front of $\langle W\rangle$
is fixed by the (effective) power-law dependence of
$\langle W\rangle$ on $y_p$.
Hence (\ref{chibar2}) is more exact
than what one might have expected from the approximate nature of
Eqs. (\ref{W-mean1}) and (\ref{W-mean2}).]

The dielectric susceptibility is finite down to
$T=0$. At $T=0$, individual pairs can not respond to a weak
applied field due to loss of entropy. 
The polarizability of the medium is a consequence of a finite density
of states at zero pair energy. Pair configurations with a slightly positive
energy in the absence of the field may acquire a negative energy
if it is favored by the field, and hence become occupied.
The opposite happens for the unfavored pair configurations
opposing the field direction.

\subsection{The pair freezing temperature $T_\ast$}

The change in the leading order behavior of the pair density 
$\rho\sim\langle W\rangle/R^2$
at $T_\ast$ has a simple interpretation.
Given the strong repulsive interaction between two pairs at a distance
smaller than their size $R$, and the absence of correlation
in the disorder potential on a pair beyond a distance of order $R$,
it is reasonable to assume that clustering of pairs is rare 
in the dilute limit.
The typical distance between neighboring pairs is thus given by
$L=\rho^{-1/2}>R$. Within an area of linear size $L$, we have
typically one pair only.

Let us first consider the equilibrium statistics of
a single pair in a box of linear size $L$, taken to be
arbitrary for the moment.
The total number of configurations available to the pair 
is $N=(R/a)^2(L/a)^2$, and the variance of the random potential,
$\Delta=4\pi\sigma J^2\ln(R/a)$.
In the random energy approximation,
the mean free energy of the pair follows from Eq. (\ref{rem-f}),
\begin{eqnarray}
\nonumber
F_p(L,T)
\simeq &&2E_c+2\pi J\ln(R/a)\\
&&-2c(T,s)[\ln(R/a)+\ln(L/a)],
\label{F-pair}
\end{eqnarray}
where 
\begin{equation}
s={2\pi\sigma J^2\ln (R/a)\over\ln(R/a)+\ln(L/a)}.
\label{s-pair}
\end{equation}
For a fixed $L$, $F_p(L,T)$ increases with decreasing $T$,
and locks to a constant for $T<T_g(s)$.
At a fixed temperature, $F_p(L,T)$ decreases with increasing $L$.

The typical inter-pair distance $L(T)$ is determined by the condition
\begin{equation}
F_p(L,T)=0.
\label{F-p=0}
\end{equation}
From the properties of $F_p$ mentioned
above, we see that $L(T)$ increases as $T$ decreases,
and locks to a constant $L_\ast$ for $T<T_\ast$. Here
$T_\ast=T_g(s_\ast)$ is obtained self-consistently,
with $s_\ast$ given by (\ref{s-pair}) at $L=L_\ast$.
The result for $T_\ast$ agrees with (\ref{T-star1}).
For $T>T_\ast$, we may use the high-temperature expression for
$c(T,s)$ in (\ref{F-pair}), and the condition (\ref{F-p=0}) yields
the following estimate for the number of pairs in an area of size $R$,
\begin{equation}
(R/L)^2\simeq (R/a)^{4-2\pi K+2\pi\sigma K^2}\exp(-2E_c/T),
\label{L(T)}
\end{equation}
in agreement with (\ref{W-mean1a}).
The length $L_\ast$ satisfies
\begin{equation}
(R/L_\ast)^2\simeq (R/a)^4\exp(-\Delta/2T_\ast^2),
\label{L-ast}
\end{equation}
in rough agreement with (\ref{W-mean2}) for the
number of pairs in an area of size $R$ below $T_\ast$.

The physical meaning of the temperature $T_\ast$
is now clear. For $T>T_\ast$, the entropy of a pair
in a region of the size of inter-pair distance
is finite and varies smoothly with $T$. 
This entropy is lost at $T_\ast$.
Therefore $T_\ast$ is associated with the pair freezing.
The length scale $L_\ast(R)$ is the smallest size of an area 
where one typically finds a negative ground state energy for a pair
of size $R$.

In contrast, the single-vortex glass temperature
$T_g$ is associated with the lost of entropy for a
pair when it is restricted to an area of pair size.
[Note that (\ref{s-pair}) reduces to the expression for
a single vortex when we set $L=R$.]
This temperature does not play a special role
in the {\it equilibrium} behavior of a pair, where the
relevant length scale is set by the inter-pair distance.
Likewise, so far as the equilibrium properties 
of a dilute gas of pairs are concerned,
the cell representation we employed is merely a convenient device 
for performing calculations.

The equivalence of our results to those of Refs. \cite{nskl} and
\cite{sche96} implies that there is a simple connection between
the two approaches. In the work of NSKL and the more recent paper
by Scheidl, calculation of thermal averages were made under 
the ``factorization-ansatz'',
which assumes that pairs do not interact
unless they take identical positions. 
From the discussion of Sec.~IV.~A we see that
the pair-pair repulsion extends to a distance of the order of
pair size $R$. If there is no strong reason provided by disorder
for clustering of pairs, the two approaches should differ only
by a relative amount proportional to the pair density, i.e.,
the difference should show up only at order
$\langle W\rangle^2$ in the expressions for $\bar\chi$, etc.
This is precisely what happens under the random-energy approximation.
In reality, due to correlations in the disorder potential,
close to a very favorable configuration for a pair, there are
other configurations which are nearly as favorable, though 
pair-pair repulsion would forbid simultaneous occupation
of these configurations. The true density of pairs is thus
expected to be somewhat smaller than the one calculated
under the factorization ansatz or the random energy approximation.
Nevertheless, from what we understand about the correlations,
the qualitative behavior of the system should be the same as
predicted by the approximate calculations.
In particular, no change in the exponent of the power-laws
in Eq. (\ref{W-mean_a}) is expected.

\section{Recursion relations and results}

\subsection{The RG transformation}

The knowledge we gained about a dilute gas
of vortex-antivortex pairs can now be incorporated into
a RG procedure aimed at capturing the large-distance behavior of
the Coulomb gas with disorder.
This can be done explicitly following an integration scheme 
used previously by Kosterlitz for treating the pure problem\cite{kost74}.

Consider a configuration $\{m_i\}$ made up of
two groups of charges. The first group,
$\{m_i^<\}$, consists of pairs of $\pm 1$ vortices,
each of size less than a cut-off size $R$. The second group,
$\{m_i^>\}$, consists of charges which do not fall into that category.
(Note that our usage of the superscripts ``$<$'' and ''$>$'' is the 
opposite of the one familiar in momentum-space RG.)
The total energy of the system, Eq. (\ref{c-gas}), can be rewritten as,
\begin{eqnarray}
\nonumber
H_{\rm v}(\{m_i\})=&&H_{\rm v}(\{m_i^<\})+H_{\rm v}(\{m_i^>\})\\
&&+H_{\rm int}(\{m_i^<\},\{m_i^>\}),
\label{H-split}
\end{eqnarray}
where
\begin{equation}
H_{\rm int}\simeq -\sum_n {\bf p}_n\cdot {\bf E}^>({\bf r}_n)
\label{H_int}
\end{equation}
describes the interaction between the two groups.
Here ${\bf p}_n$ is the dipole moment of pair $n$ in the first group,
${\bf r}_n$ is the center position of the pair, and
\begin{equation}
{\bf E}^>({\bf r})=2\pi J\sum_i m_i^>
{{\bf r}-{\bf r}_i\over |{\bf r}-{\bf r}_i|^2}
\label{E-great}
\end{equation}
is the electric field at {\bf r} due to the presently unpaired charges
in the second group.

The partition sum over the paired charges is given by
\begin{equation}
\Xi^<=\sum_{\{m_i^<\}}
\exp\bigl(-[H_{\rm v}(\{m_i^<\})+H_{\rm int}]/T\bigr).
\label{Xi_less}
\end{equation}
Writing $\Xi^<\equiv\Xi_0^<\exp(-\delta H/T)$, where
$\Xi_0^<$ is the partition function at $H_{\rm int}=0$, we obtain,
\begin{equation}
\delta H(\{m_i^>\})=-T\ln\langle \exp(-H_{\rm int}/T)\rangle_0,
\label{delta-H0}
\end{equation}
where $\langle\cdot\rangle_0$ denotes thermal average with respect
to $H_{\rm v}(\{m_i^<\})$.
Treating $H_{\rm int}$ as a perturbation,
we can write $\delta H$ in a more suggestive form,
\begin{equation}
\delta H=-\int d^2 r [{\bf P}_0({\bf r})+{1\over 2}{\bf P}_{\rm ind}({\bf r})]
\cdot {\bf E}^>({\bf r})+O(|{\bf E}^>|^3).
\label{delta-H}
\end{equation}
Here 
${\bf P}_0({\bf r})=\langle\sum_n {\bf p}_n\delta({\bf r}-{\bf r}_n)\rangle_0$
is the zero-field polarization of the paired charges in the absence of the 
interaction term $H_{\rm int}$,
and ${\bf P}_{\rm ind}={\bf P}-{\bf P}_0$ is the
induced polarization of the paired charges due to the field ${\bf E}^>$.
[Note that ${\bf P}({\bf r})$ is defined in the same way as 
${\bf P}_0({\bf r})$ except that the thermal averaging is taken
with respect to $H_{\rm v}(\{m_i^<\})+H_{\rm int}$.]

The renormalization group idea is to take the cut-off
size $R$ as a running parameter, and perform the
elimination of paired charges $\{m_i^<\}$ in a step by step manner, 
so that each time one needs to
deal with pairs in a narrow size range $R$ to $R+dR$ only.
The necessary calculations have already been done in Sec.~IV.
Substituting Eq. (\ref{chi}) into (\ref{delta-H}), we obtain,
\begin{equation}
\delta H
\simeq -\int d^2 r[{\bf P}_0\cdot{\bf E}^>+{1\over 2}\bar\chi|{\bf E}^>|^2].
\label{deltaH2}
\end{equation}
This is nothing but the field integral version of the Coulomb 
energy (\ref{c-gas}),
and hence can be incorporated into $H_{\rm v}(\{m_i^>\})$ 
by redefining the parameters $J$ and $\sigma$ of the model.

The change in $J$ can be obtained with the help of
the first example in Appendix~A. One thing to note is that
the integral over $|{\bf E}^>|^2$ in (\ref{deltaH2}) 
exclude regions of size $R$ around each $m_i^>$ charge,
since paired vortices should not be found in these areas.
This leads to the identification $b=R$ in Eq. (\ref{cal-E}).
The new effective parameters are given by,
\begin{equation}
E_c\rightarrow \tilde E_c=E_c+4\pi^3\bar\chi J\tilde J\ln(R/a),
\label{delta-Ec}
\end{equation}
and
\begin{equation}
J\rightarrow \tilde J^{-1}=J^{-1}+4\pi^2\bar\chi,
\label{delta-J}
\end{equation}
equivalent to Eq. (\ref{epsilon-new})\cite{note3}.
The extra term in $\tilde E_c$ merely accounts for the fact
that screening from this group of pairs is effective
only at distances larger than $R$.

In Sec.~IV.~C, contribution to the 
zero-field dipole moment ${\bf p}_0$ of a cell due to
disorder within the cell was calculated.
More distant disorder contributes to ${\bf p}_0$ by acting as
an additional polarizing field.
When the latter contribution is
substituted into Eq. (\ref{deltaH2}), 
we see that, with the help of the second example in Appendix~A,
the interaction strength
between $\{m_i^>\}$ and quenched dipoles ${\bf q}_{j}$
is reduced by a factor $1-2\pi^2\bar\chi J$.
Combining the zero-field polarization ${\bf P}_0={\bf p}_0/R^2$ of pairs with
the disorder polarization ${\bf Q}={\bf q}/R^2$,
we obtain the effective disorder that couples linearly to ${\bf E}^>$
in the Hamiltonian $H_{\rm v}+\delta H$ for the $\{m_i^>\}$ charges,
\begin{equation}
{\bf Q}_{\rm eff}=(1-4\pi^2\bar\chi J){\bf Q}+\tilde{\bf P}_0,
\label{tilde-Q}
\end{equation}
where $\tilde{\bf P}_0=\tilde{\bf p}_0/R^2$ is independent of ${\bf Q}$
[see discussion around Eq. (\ref{p0-split})].

Equation (\ref{tilde-Q}) shows that pair excitations modify
the quenched disorder seen by $\{m_i^>\}$ charges in two different ways. 
The first effect is the screening of the interaction
at distances larger than the pair size $R$, 
which can be taken into account by a redefinition of $J$,
Eq. (\ref{delta-J}).
The second effect is the generation of additional
disorder. Since $\tilde{\bf P}_0$ is independent of {\bf Q},
we obtain an additive contribution to the variance of disorder, $\sigma$.
Writing $\tilde J\tilde{\bf Q}=J{\bf Q}_{\rm eff}$, and
using $\langle|{\bf Q}|^2\rangle\equiv \sigma/(2\pi^2R^2)$,
we get
\begin{equation}
\tilde\sigma=\sigma+2\pi^2\langle W^2\rangle +O(\langle W\rangle^2).
\label{tilde-sigma}
\end{equation}
In deriving the above expression we used the fact that
the difference between the variance of ${\bf P}_0$ and that 
of $\tilde{\bf P}_0$ is of order $\langle W\rangle^2$.
Using the result for $\langle W^2\rangle$,
we see that the change in $\sigma$ is proportional to $\langle W\rangle$
for $T<T_\ast$, but of higher order for $T>T_\ast$.

Equations (\ref{delta-J}) and (\ref{tilde-sigma}) can be
expressed in the usual differential form by writing $R=ae^l$.
For convenience, we introduce a dimensionless
quantity $Y(R)$, such that $2\pi Y^2dR/R\equiv\langle W\rangle$ gives
the number of vortex-antivortex pairs of size between 
$R$ and $R+dR$, in an area of size $R^2$ and averaged over the whole system.
For $T>T_\ast$ or $K^{-1}\equiv T/J>K_\ast^{-1}=2\sigma$,
we have,
\begin{mathletters}
\begin{eqnarray}
\label{dk-high}
dK^{-1}/dl&=&4\pi^3Y^2,\\
\label{dsigma-high}
d\sigma/dl&=&0,\\
\label{dY-high}
dY/dl&=&(2-\pi K+\pi\sigma K^2)Y.
\end{eqnarray}
\label{flow-high}
\end{mathletters}
For $T<T_\ast$ or $K^{-1}<K_\ast^{-1}$, we have,
\begin{mathletters}
\begin{eqnarray}
\label{dk-low}
dK^{-1}/dl&=&2\pi^3\sigma^{-1}K^{-1}Y^2,\\
\label{dsigma-low}
d\sigma/dl&=&2\pi^3(2-\sigma^{-1}K^{-1})Y^2,\\
\label{dY-low}
dY/dl&=&\Bigl(2-{\pi\over 4\sigma}\Bigr)Y.
\end{eqnarray}
\label{flow-low}
Note that, in writing the above equations, we only kept
terms up to order $Y^2$.
The flow equations for $Y$ follow from the power-law dependence
of the pair density on pair size as given by
Eqs. (\ref{W-mean1a}) and (\ref{W-mean2a}).
A term of order $1/l$ inside the brackets in (\ref{dY-low}) has
been neglected.
\end{mathletters}

A few remarks concerning the above recursion relations are in order.
For $T>T_\ast$, Eqs. (\ref{flow-high}) is identical to
those of previous authors\cite{rubi83,card82}. 
From Eq. (\ref{tilde-sigma}) we
see that there is a renormalization of $\sigma$ even in
this regime, but the effect is of higher order than $Y^2$.
The change of the flow equations for $T<T_\ast$ was pointed
out earlier by NSKL\cite{nskl}. The renormalization of $\sigma$, though
not recorded previously, has also been obtained by Scheidl 
using a different approach\cite{sche96}.

In the absence of disorder, $Y$ is equal to
the ``rescaled'' single vortex fugacity, $(R/a)^{2-\pi J/T}\exp(-E_c/T)$,
in the dilute limit.
When disorder is present,
relation between $Y$ and the core energy $E_c$ is
more complicated.
The bare value of $Y$ can be obtained from Eqs. (\ref{W-mean1a}) 
and (\ref{W-mean2a}) for $T>T_\ast$ and $T<T_\ast$, respectively.
It has a finite limit $Y_0$ at $T=0$. For small $\sigma$, 
$Y_0^2\sim \sigma^{1/2}\exp(-c/\sigma)$, where $c$ is a positive,
model-dependent number.
At small values of $T$, the bare value of $Y$ increases from $Y_0$
by an amount proportional to $T^2$.

\subsection{Phase diagram and thermodynamic properties}

\subsubsection{Constants of RG flow}

Seemingly complex at first sight, the
flow equations (\ref{flow-high}) and (\ref{flow-low}) have in fact the
same structure as their $\sigma=0$ counterpart\cite{kost74}.
The fixed points of the flow are located on the $Y=0$ plane
in the three-dimensional (3D) parameter space 
spanned by $K^{-1}$, $\sigma$, and $Y$.
They are stable in the region enclosed by the dashed line
in Fig. 4, but unstable outside the region.
Points on the dashed line are hyperbolic fixed points
which describe the pair-unbinding transition.

It turns out that the flow equations are completely integrable.
In the region $T>T_\ast$ [to the right of the dotted line
in Fig. 4], 
\begin{mathletters}
\label{1st}
\begin{equation}
\sigma ={\rm const}.
\label{1st-high}
\end{equation}
is obviously a constant of the flow. 
On the $T<T_\ast$ side, the corresponding
first integral is given by
\begin{equation}
K-\sigma K^2={\rm const.}
\label{1st-low}
\end{equation}
as can be easily verified using Eqs. (\ref{dk-low}) and (\ref{dsigma-low}).
These ``streamlines'' of the flow are illustrated in Fig. 4.
\end{mathletters}

The second constant of the flow is given by 
\begin{mathletters}
\label{2nd}
\begin{equation}
Y^2-\pi^{-3}(K^{-1}-{\pi\over 2}\sigma K+{\pi\over 2}\ln K)={\rm const.}
\label{2nd-high}
\end{equation}
for $T>T_\ast$, and
\begin{equation}
Y^2-\pi^{-3}(K^{-1}+\sigma+{\pi\over 4}\ln K)={\rm const.}
\label{2nd-low}
\end{equation}
for $T<T_\ast$. A unique trajectory in the 3D parameter space
is specified when one combines (\ref{2nd}) with (\ref{1st}),
with the constants fixed by bare values of the parameters involved.
\end{mathletters}

\subsubsection{Phase diagram}

The original XY model has only two parameters, $K_B^{-1}=T/J$ and
$\sigma_B$. The bare value of $Y$, $Y_B$,
is a function of $K_B$ and $\sigma_B$.
The phase boundary of the model is determined by the condition
that the RG flow ends on the dashed line in Fig. 4.
Since the flow takes both $\sigma$ and $K^{-1}$ to larger values,
this phase boundary lies within the area bounded by the dashed line,
as illustrated in Fig. 5.

The question of reentrance of the disordered phase is
whether the upper-left part of the phase boundary illustrated in Fig. 5
contains a piece with a positive slope.
Although earlier calculations which led to
the prediction of a reentrance transition are not to be trusted,
it is actually difficult to rule out such a possibility from
the new RG flow equations (\ref{flow-low}).
From Eq. (\ref{dsigma-low}) we see that, at a given $Y^2$,
the increase in the effective disorder becomes slower 
as temperature increases.
Hence for a fixed $Y_B$, the bare value of $\sigma$
which flows to the fixed point value $\sigma=\pi/8$ increases
with increasing $T/J$. 
On the other hand, in the original XY model, the bare
pair density $Y_B^2$ at a fixed $\sigma$ is expected to 
increase with temperature, too.
The calculation presented in this paper is not
quantitative enough to assess the two competing effects
to reach a precise conclusion on the shape of the low temperature part of
the phase boundary. 

\subsubsection{Approach to criticality}

The critical behavior around the transition is controlled by the
RG flow close to the relevant hyperbolic fixed point.
As an example, let us consider flows along a particular contour
in Fig. 4. Substituting Eqs. (\ref{1st}) into Eqs. (\ref{2nd}),
we get a flow pattern depicted in Fig. 6.
The curve consists of two pieces, one from Eqs. (\ref{1st-low}) and 
(\ref{2nd-low}) for $K^{-1}<K^{-1}_\ast$, and the other 
from Eqs. (\ref{1st-high}) and (\ref{2nd-high}) for $K^{-1}>K_\ast^{-1}$.
Since Eqs. (\ref{1st}) do not involve $Y^2$,
any vertical translation of the curve shown in Fig. 6 is
also an invariant of the RG flow.
This family of curves can be parametrized by the minimum value of $Y^2$ 
on each curve, $D$.

In the ordered phase and at the transition, the scaled
pair density $Y^2$ eventually reaches zero on large
length scales. The corresponding RG flow follows a curve
with $D\leq 0$, with the marginal case $D=0$ reserved for the transition.
On the disordered side but close to the transition,
the RG flow follows a curve with a small positive $D$.
Along such a curve, the value of $Y^2$ first decreases as the
running parameter $l$ increases,
becomes almost stationary around the minimum of the curve, 
and then grows rapidly to large values beyond $l=l_+$.
The length $\xi_+=a\exp(l_+)$ sets the scale where the
correlation function (see Sec.~V.~C below) turns from 
a power-law to an exponential decay.

The dependence of $\xi_+$ on $D$ can be obtained by examining
the flow in the vicinity of the minimum of the curve at $K^{-1}=K^{-1}_f$
(see Fig. 6).
Let $X=K^{-1}-K_f^{-1}$ be the distance from the minimum. 
Generically, for small $X$, we have
\begin{equation}
Y^2=\lambda X^2+D,
\label{minimum-1}
\end{equation}
where $\lambda$ is proportional to the curvature at the minimum.
It is easy to check that, in this case, the coefficient
of $Y$ on the right-hand-side of either Eq. (\ref{dY-high})
or (\ref{dY-low}), which ever applies,
is linearly proportional to $X$.
Using (\ref{minimum-1}), we can write the flow equation for $Y$ as,
\begin{equation}
dY/dl=\gamma^{-1}(Y^2-D)^{1/2}Y,
\label{dY-a1}
\end{equation}
where $\gamma$ is a nonuniversal number which depends on the 
location of the fixed point.
The parameters $D$ and $\gamma$ can be scaled away using the substitution
$Y\rightarrow D^{1/2}Y$ and $l\rightarrow \gamma D^{-1/2}l$.
This yields a correlation length
\begin{equation}
\xi_+=a\exp(l_+)\simeq a\exp(\gamma/D^{1/2}).
\label{xi-c}
\end{equation}
The dependence of $D$ on the shift of the bare parameters 
from their critical values can be obtained by 
solving (\ref{1st}) and (\ref{2nd}).

There is, however, a special case where the curvature of the
curve at the minimum vanishes, invalidating the above analysis.
This happens at the point $S$ (and correspondingly $S'$)
in Fig. 5, where the $T_\ast$-line meets the phase boundary.
The minimum of the curve shown in Fig. 6 is now at the meeting
point of the high and low temperature segments.
This is also the inflection point of each of the two curves.
The function around the minimum takes the form,
\begin{equation}
Y^2=\tilde\lambda |X|^3+D,
\label{minimum-2}
\end{equation}
where $\tilde\lambda$ is another constant.
The flow rate of $Y$ is now proportional to $X^2Y$.
Consequently, we have,
\begin{equation}
dY/dl=\tilde\gamma^{-1}(Y^2-D)^{2/3}Y,
\label{dY-a2}
\end{equation}
where $\tilde\gamma$ is yet another number.
The proper scaling in this case is
$Y\rightarrow D^{1/2}Y$ and $l\rightarrow \tilde\gamma D^{-2/3}l$.
A new dependence of the correlation length on $D$ follows,
\begin{equation}
\xi_+\simeq a\exp(\tilde\gamma/D^{2/3}).
\label{xi-c2}
\end{equation}

\subsubsection{Free energy}

Let us now turn to the behavior of the free energy in the 
vicinity of the transition.
Following the discussion of Sec.~IV.~A, we may write 
the contribution to the free energy per unit area 
from pairs in the size range $R$ to $R+dR$ as,
\begin{equation}
dF_{\rm v}\simeq -T\langle\ln(1+y_pz_p)\rangle/R^2.
\label{F-v}
\end{equation}
Comparing the above equation with (\ref{W}), we obtain,
\begin{equation}
y_p{\partial (dF_{\rm v})\over\partial y_p}=-T\langle W\rangle/R^2.
\label{F-v-r}
\end{equation}
By making analogy to Eqs. (\ref{chibar1}) and (\ref{chibar2}),
and using the definition of $Y$, we find,
\begin{equation}
{dF_{\rm v}\over d\ln R}\simeq
\left\{
\begin{array}{l}
-2\pi T Y^2/R^2,\;\;\;\mbox{for}\; T>T_\ast;\\
\\
-2\pi T_\ast Y^2/R^2,
\;\;\mbox{for}\; T<T_\ast.
\end{array}\right.
\label{df-v-f} 
\end{equation}

The total vortex contribution $F_{\rm v}$ to the free energy density of the
XY model is obtained by integrating $dF_{\rm v}$ over $R$.
Let $F_{{\rm v},c}$ be the vortex free energy density at the transition.
Equation (\ref{df-v-f}) then yields,
\begin{equation}
F_{\rm v}-F_{{\rm v},c}\sim -\int_a^\infty R^{-3}dR (Y^2-Y_c^2),
\label{delta-F-v}
\end{equation}
where $Y_c(R)$ is the value of $Y(R)$ at the transition.
A full analysis of the integral is quite involved, but
the following consideration should yield a correct estimate
of the expected singular behavior.

Approaching the transition from the ordered phase,
$Y$ flows to $0$ at $R=\infty$. 
The difference $Y^2(R)-Y_c^2(R)$ is expected to remain constant,
say equal to $D<0$,
up to $R=\xi_-$, and then goes to 0.
Truncating the integral at $R=\xi_-$, we obtain,
\begin{equation}
F_{{\rm v},-}-F_{{\rm v},c}\sim -Da^{-2}+D\xi_-^{-2}.
\label{F-sin-low}
\end{equation}
The crossover length $\xi_-$ has the same behavior as $\xi_+$,
except that one should replace $D$ by $-D$ in Eqs. (\ref{xi-c})
and (\ref{xi-c2}), as the case may be.
From the disordered side, the story is the same up to $R=\xi_+$,
but beyond that $Y^2$ becomes of order 1. Hence we expect
\begin{equation}
F_{{\rm v},+}-F_{{\rm v},c}\sim -Da^{-2}-\xi_+^{-2}.
\label{F-sin-high}
\end{equation}
The singularity of $F_{{\rm v},\pm}$ are thus related to
the singular behavior of $\xi_\pm$.
In both cases, it is an essential singularity.

Finally, it is interesting to see if the pair-freezing line
$T_\ast$ in the ordered phase corresponds to another
singularity of the vortex free energy.
The analysis presented in Appendix~D suggests that this is not
the case. We should however note that
the absence of a true glass transition in our model is a result of
the very special type of functional dependence of
relevant quantities on temperature and pair size,
and hence might be susceptible to various types of perturbations.

\subsection{Two-point correlation function}

Consider now the two-point phase-phase correlation function,
\begin{equation}
C(r_{ij})\equiv\bigl\langle\exp\bigl[i(\phi_i-\phi_j)\bigr]\bigr\rangle,
\label{C-full}
\end{equation}
where the average is taken over thermal and then disorder fluctuations.
When only spin-wave contributions to $\phi$ are taken into account,
one obtains a power-law decay of $C$ with distance $r_{ij}$, 
as shown in Eq. (\ref{2pt}).
Vortex-pair excitations ``soften'' the spin-waves,
and lead to a faster decay of the correlation function.
On grounds of renormalizability of the model, 
one expects that, in the ordered phase,
\begin{equation}
C(r)\sim r^{-\eta}
\label{C-full-pl}
\end{equation}
at large distances, but the exponent $\eta$ differs
from $\eta_{\rm sw}$ in that the renormalized values 
$K^{-1}_\infty$ and $\sigma_\infty$ should be used,
\begin{equation}
\eta={1\over 2\pi}(K_\infty^{-1}+\sigma_\infty).
\label{eta-inf}
\end{equation}

The above conjecture for the average behavior of $C(r)$
can be verified by a direct calculation.
Since the spin-wave and vortex fluctuations decouple, and
the disorder which enters $H_{\rm sw}$ is orthogonal to
the disorder in $H_{\rm v}$ (see Sec.~II.~A),
we may write,
\begin{equation}
C(r)=C_{\rm sw}(r)C_{\rm v}(r),
\label{C-factorize}
\end{equation}
where $C_{\rm v}(r)$ is the correlation of phases $\phi_{\rm v}$
generated by vortex excitations,
\begin{equation}
C_{\rm v}(r_{ij})\equiv
\bigl\langle\exp\bigl[i(\phi_{{\rm v},i}-\phi_{{\rm v},j})\bigr]\bigr\rangle.
\label{C-vortex}
\end{equation}

Following Kosterlitz\cite{kost74}, we calculate the right-hand-side of
(\ref{C-vortex}) by successive elimination of
vortex-antivortex pairs, starting from the smallest pair size.
The formula below gives the difference in the phase at two sites $i$ and $j$,
generated by a single vortex-antivortex pair at ${\bf r}_p$,
\begin{equation}
\phi_{{\rm v},i}-\phi_{{\rm v},j}\simeq
(2\pi J)^{-1}({\bf p}\times\hat {\bf z})\cdot {\bf E}({\bf r}_p),
\label{phase-diff}
\end{equation}
where ${\bf p}$ is the dipole moment of the pair
(assumed to have a magnitude $R$ much smaller than $r$) 
and ${\bf E}({\bf r}_p)$ is the electric field at ${\bf r}_p$
due to a $+1$ charge placed at site $i$ and a $-1$ charge placed
at site $j$.
The variance of the phase difference
generated by {\it all pairs} in the size range $R$ to $R+dR$ can
now be easily calculated with the help of the equivalence of
Eqs. (\ref{field-e}) and (\ref{cal-E}),
\begin{equation}
\langle(\phi_{{\rm v},i}-\phi_{{\rm v},j})^2\rangle_{dR}
\simeq 2\pi\langle W\rangle\ln (r_{ij}/R).
\label{dphi-2}
\end{equation}
Going from $R$ to $R+dR$, $C_{\rm v}(r)$ is reduced by a factor
\begin{equation}
\exp\bigl[-{1\over 2}
\langle(\phi_{{\rm v},i}-\phi_{{\rm v},j})^2\rangle_{dR}\bigr]
\simeq (r/R)^{-d\eta}.
\label{increment}
\end{equation}
Comparing the coefficient of the logarithm in (\ref{dphi-2})
with Eqs. (\ref{p0-var}), (\ref{chibar}), and (\ref{tilde-sigma}),
we obtain,
\begin{equation}
d\eta={1\over 2\pi}(dK^{-1}+d\sigma),
\label{deta}
\end{equation}
which is nothing but the differential form of (\ref{eta}).
This confirms the intuitive idea that the asymptotic value
of $\eta$ is given by (\ref{eta-inf}).
On the transition line, $\eta$ decreases monotonically 
from the value $1/4$ for
the pure case to $1/16$ at the zero-temperature transition point\cite{nskl}.

Inspecting Eq. (\ref{phase-diff}), we see that
spatial and angular localization of the vortex-antivortex pairs
may lead to rare, but significant deviation of 
the thermally averaged correlation
function at two fixed sites $i$ and $j$,
\begin{equation}
\tilde C_{\rm v}({\bf r}_i,{\bf r}_j)\equiv
\bigl\langle\exp\bigl[i(\phi_{{\rm v},i}-\phi_{{\rm v},j})\bigr]
\bigr\rangle_{\rm thermal},
\label{C-therm}
\end{equation}
from its average value, $C_{\rm v}(r_{ij})$.
Such a fluctuation comes about when we are looking at large-size
pairs which have strong density fluctuations (after thermal averaging)
on the scale $r_{ij}$,
and that the presence or absence of such a pair in the region
surrounding $i$ and $j$ will 
make a significant change to the phase difference 
$\phi_{{\rm v},i}-\phi_{{\rm v},j}$.
Below $T_\ast$, the typical density of these pairs 
is significantly smaller than the average density.
Consequently, the typical value of the exponent,
$\eta_{\rm typ}$, can be somewhat smaller than its average value $\eta$.

\section{Summary}

The main conclusions of the present work can be summarized as follows.
When the variance $\sigma$ of the random phase shifts is
smaller than a critical value $\sigma_c$,
the quasi-long-range order of the 2D XY model survives at
sufficiently low temperatures. The value of $\sigma_c$
is nonuniversal but should be smaller than $\pi/8$.
The two-point phase-phase correlation decays algebraically
with distance in the entire ordered phase up to the transition.
At the transition, the exponent of the power-law decay
lies in the range between $1/16$ and $1/4$.
Approaching the transition from the disordered side,
the correlation length diverges exponentially
with a $-{1\over 2}$- or $-{2\over 3}$-power of the 
distance from the transition point.
The free energy exhibits an essential singularity on the phase
boundary.
The low temperature region of the disordered phase at
$\sigma>\sigma_c$ is not expected to show glassy long-range order.

The behavior of vortex-antivortex pairs inside the ordered phase is
quite interesting. In contrast to the pure case, there is a finite
density of such pairs at zero temperature. 
As the temperature $T$ increases,
the pair density increases initially slowly up to 
$T=T_\ast$, and then grows rapidly as entropy comes into play.
Following an opposite sequence, 
as $T$ is lowered from the transition temperature, 
large-size pairs undergo various
degrees of localization both in space and in its angular distribution.
However, in the Coulomb gas language, a finite susceptibility
for the gas of pairs is found at all temperatures.
Localization also introduces a zero-field random polarization
of the gas of pairs, which has the effect of enhancing disorder
seen by large-size pairs.

Much of the qualitative aspects of our results agree 
with those of NSKL\cite{nskl} and of Cha and Fertig\cite{cha95},
though there are minor but important differences on a quantitative level.
Technically, the analogy introduced here to the random energy model
offers a simple way to understand some of the subtle features
introduced by the disorder potential, responsible for
the failure of previous calculations based on a small-fugacity
expansion. A somewhat surprising result is that no singularity
in the free energy is found at $T_\ast$.
In fact, we have shown that an expansion with respect to
the pair-fugacity $y_p$ fails for any $\sigma>0$.
The whole analytic structure of the free energy as a function
of $T$ and $\sigma$ remains to be explored.

A new quantity which appeared in this paper is the 
single-vortex glass temperature $T_g$.
This temperature also signals localization in the 
angular distribution of a pair when translation of the pair
over a distance larger than its size is forbidden.
As we have seen, this temperature plays no special role
in the thermodynamics of the XY model. Nevertheless, one
might contemplate possible changes of dynamical behavior
at $T_g$, an issue to be studied further.

\section*{Acknowledgements}

During the course of this work,
I have benefitted greatly from enlightening discussions
with Bernard Derrida, Sergei Korshunov, Thomas Nattermann, and 
Stephan Scheidl.
Part of the work was done at the Weizmann Institute
during a very pleasant workshop organized by 
Profs. David Mukamel and Eytan Domany.
Research is supported in part by
the German Science Foundation through project SFB-341.

\appendix
\section{Electrostatics in two dimensions}

In this Appendix~we collect some useful formula
from electrostatics in two dimensions.
We adopt the convention that the unit of charge is $1$.
The electric potential due to a $+1$ charge at origin is given by
\begin{equation}
V({\bf r})=-\epsilon^{-1}\ln (|{\bf r}|/a)
\label{V-potential}
\end{equation}
where $\epsilon=(2\pi J)^{-1}$ is the bare dielectric constant.
The electric field ${\bf E}=-\nabla V$ satisfies
\begin{equation}
\nabla\cdot{\bf E}=2\pi\epsilon^{-1}\rho({\bf r}),
\label{Poisson}
\end{equation}
where $\rho({\bf r})$ is the charge density at {\bf r}.

In a dielectric medium, the induced polarization ${\bf P}=\chi\;{\bf E}$,
where $\chi$ is known as the dielectric susceptibility.
The displacement vector ${\bf D}=\epsilon{\bf E}+2\pi{\bf P}$
satisfies
\begin{equation}
\nabla\cdot{\bf D}=2\pi\rho_f({\bf r}),
\label{Poisson1}
\end{equation}
where $\rho_f({\bf r})$ is the density of free charges at {\bf r}.
Writing ${\bf D}=\tilde\epsilon\;{\bf E}$, we obtain
\begin{equation}
\tilde\epsilon=\epsilon+2\pi\chi.
\label{epsilon-new}
\end{equation}

Let us now consider three examples encountered in the main text.
In the first example we establish the equivalence of two expressions
for the energy of a charge-neutral system.
The electric field generated by a set of point charges
$m_i$ located at ${\bf r}_i$ is given by
\begin{equation}
{\bf E}({\bf r})=2\pi J\sum_i m_i
{{\bf r}-{\bf r}_i\over |{\bf r}-{\bf r}_i|^2}.
\label{E-field}
\end{equation}
Consider now the integral
\begin{equation}
{\cal E}={1\over 8\pi^2 J}\int d^2r\ |{\bf E}|^2.
\label{field-e}
\end{equation}
Writing 
\begin{equation}
|{\bf E}|^2=V\nabla\cdot{\bf E}-\nabla\cdot (V{\bf E}),
\label{separate}
\end{equation}
and using the Gauss's theorem and Eq. (\ref{Poisson}),
we obtain,
\begin{equation}
{\cal E}=\sum_i m_i^2E_c-\pi J\sum_{i\neq j}m_im_j\ln(|{\bf r}_i-{\bf r}_j|/b).
\label{cal-E}
\end{equation}
Here
\begin{equation}
E_c={J\over 2}\int_{|{\bf r}|<b} d^2 r |{\bf r}|^{-2}
\label{core-e}
\end{equation}
is the ``core energy'' of a unit charge. [The divergence of
(\ref{core-e}) at small distances is cut off by the existence of a lattice.]
Note that, in continuum, $b$ is an arbitrary parameter 
which does not influence the result (\ref{cal-E}).

As the second example we consider an expression for the interaction 
energy between a set of charges $m_i$ at ${\bf r}_i$ and 
quenched dipoles ${\bf q}_{\; j}$ at ${\bf R}_j$. 
Let ${\bf E}({\bf r})$ be the electric field generated
by all the charges, and ${\bf E}_d({\bf r})$ be the field due to the
dipoles but excluding those within a distance $R$ from {\bf r}.
Consider now the integral
\begin{equation}
{\cal E}_1={1\over 4\pi^2 J}\int
d^2 r\ {\bf E}\cdot{\bf E}_d.
\label{cal-E1}
\end{equation}
We now write ${\bf E}_d$ as a sum over the field due to
individual dipoles.
Using again (\ref{separate}) and the Gauss's theorem separately
for each $j$, we obtain,
\begin{equation}
{\cal E}_1=\sum_i m_iV_d({\bf r}_i)+\sum_j U_j,
\label{cal-E1a}
\end{equation}
Here $V_d({\bf r})$ is the potential due to dipoles
outside a circle of radius $R$ centered at {\bf r},
and
\begin{equation}
U_j={1\over 2\pi}\int_0^{2\pi}d\phi\ {\bf R}\cdot{\bf E}({\bf R}_j+{\bf R})
{{\bf q}_{\; j}\cdot{\bf R}\over R^2},
\label{I-j}
\end{equation}
where ${\bf R}=R(\cos\phi,\sin\phi)$.
Elementary calculation yields 
$U_j={1\over 2}{\bf q}_{\; j}\cdot{\bf E}({\bf R}_j)$.
It follows that the second sum on the right-hand-side of (\ref{cal-E1a})
is $-{1\over 2}$ times the first sum. Hence
\begin{equation}
{\cal E}_1={1\over 2}\sum_i m_iV_d({\bf r}_i).
\label{cal-E1b}
\end{equation}

Our final example concerns the field ${\bf E}_Q$ inside a circle
of radius $R$ generated by a medium with a permanent polarization
{\bf Q} which fills the circle.
To by-pass an explicit calculation, we use the result that, in a
uniform external field ${\bf E}_0$,
the field inside such a circle filled with a {\it polarizable} medium
of dielectric constant $\epsilon_1$ is given by
\begin{equation}
{\bf E}_1={2\epsilon\over\epsilon+\epsilon_1}{\bf E}_0,
\label{polar-E-1}
\end{equation}
where $\epsilon$ is the dielectric constant of the medium outside the circle.
The polarization inside the circle is 
${\bf Q}=(\epsilon_1-\epsilon){\bf E}_1/(2\pi)$, and the
the field it produces is ${\bf E}_Q={\bf E}_1-{\bf E}_0$.
Using the above results, we obtain,
\begin{equation}
{\bf E}_Q=-(\pi/\epsilon){\bf Q}=-2\pi^2J{\bf Q},
\label{E-Q}
\end{equation}
which is uniform inside the circle.

\section{Freezing and long-tails in the random energy model}

As shown by Derrida\cite{derr81}, 
the freezing transition in the random energy model can be 
understood as a switching of terms which contribute most to the partition sum
(\ref{rem-partition}).
For $T<T_g$, the lowest of the $N$ energies $x_i$ dominates $z$,
while for $T>T_g$, typically a finite fraction of the $N$ energies contribute
significantly to $z$.
(The word ``typical'' refers to events which occur with a large probability,
and ``rare'' refers to events with a very small probability.)
This can be seen more explicitly as follows.

In a given realization of the disorder,
the random energies fall into a band
$x_{\rm min}\leq x_i\leq x_{\rm max}$, where  
$x_{\rm min}=\min_i\{x_i\}$ and
$x_{\rm max}=\max_i\{x_i\}$.
Introducing the integrated density of state,
${\cal N}(x)$, which gives the number of levels with $x_i\leq x$,
one can rewrite Eq. (\ref{rem-partition}) as 
\begin{equation}
z=\exp(-x_{\rm min}/T)+\int_{x_{\rm min}}^\infty \exp(-x/T)d{\cal N}.
\label{partition1}
\end{equation}
The contribution from the lowest energy level
has been isolated from the rest.
When the total number of levels is large, there are typically
many levels in an interval $dx\sim T$, so that one may replace
$d{\cal N}$ by its mean,
\begin{equation}
d{\cal N}\simeq N\psi(x)dx.
\label{anneal}
\end{equation}
Equation (\ref{anneal}) fails when ${\cal N}(x)<1$.
This happens for $x<x_0$, where $x_0$ is determined by
the condition ${\cal N}(x_0)=1$.
From Eqs. (\ref{anneal}) and (\ref{rem}) we get
\begin{equation}
x_0=-(2s)^{1/2}\ln N+O(\ln\ln N).
\label{v0}
\end{equation}
Substituting (\ref{anneal}) into (\ref{partition1}), and
restricting the integral to $x>x_0$, we obtain
\begin{equation}
z\simeq \exp(-x_{\rm min}/T)+z_{\rm typ}
\label{partition2}
\end{equation}
where
\begin{equation}
z_{\rm typ}=N\int_{x_0}^\infty dx\psi(x)\exp(-x/T).
\label{ztyp}
\end{equation}

It is easy to show that the typical value of $x_{\rm min}$
is given by $x_0$, while the typical value of
the partition function is given by 
\begin{equation}
z_{\rm typ}\simeq \exp(-\langle f\rangle/T).
\label{z-typ1}
\end{equation}
Thus the contribution from the lowest energy level to $z$ becomes
significant in a {\it typical} realization of the disorder
only when $T\leq T_g$.

On the other hand, fluctuations of $z$ far away from $z_{\rm typ}$
(say $z>2z_{\rm typ}$) are dominated by fluctuations of $x_{\rm min}$
at all temperatures.
This is especially so for $T<T_g$, where the fluctuations of $z$
are typically of order $z_{\rm typ}$.
To characterize this behavior more precisely, let us
consider the distribution of $x_{\rm min}$, denoted by 
$\psi_{\rm min}(x)$.
For the minimum of the $N$ energies 
to be greater than a certain number $x$, all $N$ energies
must be greater than $x$. Hence we have
\begin{equation}
\int_x^\infty dy\psi_{\rm min}(y)=\bigl[\int_x^\infty \psi(y)dy\bigr]^N.
\label{prob-min}
\end{equation}
For $x\ll -\sqrt{\Delta}$, the integral on the right-hand-side 
is very close to one, in which case one can write,
\begin{equation}
\psi_{\rm min}(x)=-{d\over dx}\exp\bigl[-N\int_{-\infty}^x\psi(y)dy\bigr].
\label{rho-min}
\end{equation}
For $x<x_0$, the argument of the exponential is less than 1, 
and hence
\begin{equation}
\psi_{\rm min}(x)\simeq N\psi(x)
={1\over \sqrt{2\pi\Delta}}
\exp\Bigl[
\bigl(1-{x^2\over x_0^2}\bigr)\ln N\Bigr].
\label{rho-minapp}
\end{equation}

The distribution of $x_{\rm min}$ gives us an idea about
the high-end of the distribution of the partition function
$z$, where we can write $z\simeq \exp(-x_{\rm min}/T)$.
On a log-log plot, the local slope of the distribution
is essentially given by
\begin{equation}
\zeta(z)={d\ln\psi_{\rm min}\over d\ln z}
\simeq {T\over T_g}{T\ln z\over x_0},
\label{zeta}
\end{equation}
which is a slow-varying function of $z$.
This implies that the distribution of $z$ has a long tail.
High moments of $z$ are sensitive to the tail of the distribution.
Using (\ref{rho-minapp}), we find,
\begin{equation}
{\langle z^n\rangle\over\langle z\rangle^n}\simeq
\left\{
\begin{array}{l}
1,\;\;\mbox{for}\; T>T_n;\\
\\
N^{(n-1)[(T_n/T)^2-1]},\;\;\mbox{for}\; T<T_n;
\end{array}\right.
\label{z-n}
\end{equation} 
where 
\begin{equation}
T_n=n^{1/2}T_g.
\label{T-n}
\end{equation}
The result agrees with an exact calculation by Derrida starting
from (\ref{rem-partition}).

\section{Mean pair density and fluctuations}

The disorder average of the cell occupation number $W$ can be calculated
by expanding Eq. (\ref{W}) as a power series of $y_pz_p$ 
for $y_pz_p<1$, and a power series of $(y_pz_p)^{-1}$ for
$y_pz_p>1$.
Denoting by $P(z)$ the probability distribution of $z_p$, 
we obtain,
\begin{equation}
\langle W\rangle=\sum_{n=1}^\infty (-1)^{n-1}I_1(n)
+\sum_{n=0}^\infty (-1)^{n}I_2(n),
\label{W-expand}
\end{equation}
where
\begin{eqnarray}
I_1(n)&=&y_p^n\int_0^{1/y_p}z^n P(z)dz,
\label{I-1}\\
I_2(n)&=&y_p^{-n}\int_{1/y_p}^\infty z^{-n} P(z)dz.
\label{I-2}
\end{eqnarray}

In principle, evaluation of the integrals $I_1$ and $I_2$ 
requires full knowledge of $P(z)$, which we do not have at hand.
On the other hand, as we discussed in Appendix~B,
the large-$z$ tail of $P(z)$ is due to fluctuations of
the minimum energy $x_{\rm min}$. Thus for large $z$
the substitution 
\begin{equation}
P(z)dz\rightarrow \psi_{\rm min}(x)dx\simeq
N\psi(x)dx, 
\label{P(z)}
\end{equation}
with $z=\exp(-x/T)$, yields a good approximation.
The integral $I_2$ is now readily calculated. The result reads,
\begin{equation}
I_2(n)\simeq {N\over 2}y_p^{-n}\exp\bigl({n^2\Delta\over 2T^2}\bigr)
\Bigl[1-{\rm erf}\Bigl(({n\over T}+{1\over T_\ast})\sqrt{\Delta\over 2}
\Bigr)\Bigr],
\label{I-2app}
\end{equation}
where ${\rm erf}(u)=2\pi^{-1/2}\int_0^u\exp(-x^2)dx$ is the error function.
Here 
\begin{equation}
T_\ast\equiv -\Delta/(T\ln y_p)
\label{T-star}
\end{equation}
which coincides with (\ref{T-star1}) in the limit $R\rightarrow\infty$.

The integral $I_1(n)$ can be written as
\begin{equation}
I_1(n)=y_p^n\langle z^n\rangle-I_2(-n).
\label{I-1a}
\end{equation}
To obtain its leading order behavior,
we need to examine which part of the distribution $P(z)$
contributes most to the average $\langle z^n\rangle$.
For $T>T_n=n^{1/2}T_g$, the main contribution comes from
the central part of $P(z)$ around $z_{p,\rm typ}$,
so that
\begin{equation}
I_1(n)\simeq (y_p\langle z_p\rangle)^n.
\label{I-1b}
\end{equation}
For $T<T_n$ and $T<nT_\ast$, the main contribution comes
from the tail of $P(z)$ at $z>y_p^{-1}$. In this case, the two terms
on the right-hand-side of (\ref{I-1a}) almost cancel each other.
The main contribution to $I_1$ thus comes from $P(z)$ around
$z=y_p^{-1}$, where again the approximate expression for the tail of $P(z)$
can be used. This yields
\begin{equation}
I_1(n)\simeq 
{N\over 2}y_p^n\exp\bigl({n^2\Delta\over 2T^2}\bigr)
\Bigl[1-{\rm erf}\Bigl(({n\over T}-{1\over T_\ast})\sqrt{\Delta\over2}
\Bigr)\Bigr].
\label{I-1c}
\end{equation}
When $T_n<nT_\ast$, the leading order behavior switches from
(\ref{I-1b}) to (\ref{I-1c}) at $T_n$, though (\ref{I-1c})
appears as a subleading order term in the temperature range
$T_n<T<nT_\ast$. (This is due to the fact that, around $T=T_n$,
$\langle z^n\rangle$ picks up significant contributions
from $z\simeq z_{p,\rm typ}$ and from the tail at $z>y_p^{-1}$.)
On the other hand, when $T_n>nT_\ast$, there is an intermediate
temperature range $nT_\ast<T<T_n$ where $I_1$ is dominated by
contributions from $z$ between $z_{p,\rm typ}$ and $y_p^{-1}$.
In this case, $I_1(n)\simeq y_p^n\langle z^n\rangle$, which
essentially coincides with the expression (\ref{I-1c}).
For $n=1$, (\ref{I-1c}) is valid at all temperatures.

A careful analysis of the above expressions is quite cumbersome,
but the following observations are useful and suffice for 
our purpose.

(i) Leading order behavior. For $T>T_\ast$, $\langle W\rangle$
is dominated by $I_1(1)$, 
\begin{equation}
\langle W\rangle\simeq Ny_p\exp(\Delta/2T^2).
\label{W-mean1}
\end{equation}
In terms of parameters of the model, we have
\begin{equation}
\langle W\rangle\simeq
2\pi\exp\bigl(-{2E_c\over T}\bigr)
\Bigl({R\over a}\Bigr)^{4-2\pi K+2\pi\sigma K^2}{dR\over R}.
\label{W-mean1a}
\end{equation}
For $T<T_\ast$,
$I_1(n)$ and $I_2(n)$ all contribute.
Using the asymptotic expression 
${\rm erf}(u)\simeq 1-\pi^{-1/2}u^{-1}\exp(-u^2)$ at large $u$,
we obtain
\begin{equation}
\langle W\rangle\simeq
A\Bigl({T\over T_\ast}\Bigr)
{NT_\ast\over(2\pi\Delta)^{1/2}}\exp(-\Delta/2T_\ast^2),
\label{W-mean2}
\end{equation}
or more explicitly,
\begin{equation}
\langle W\rangle\simeq
A\Bigl({T\over T_\ast}\Bigr)\Bigl({2\sigma\over\ln(R/a)}\Bigr)^{1/2}
\Bigl({R\over a}\Bigr)^{4-\pi/(2\sigma)}{dR\over R}.
\label{W-mean2a}
\end{equation}
Here 
\begin{equation}
A(u)=u\sum_{n=-\infty}^\infty{(-1)^n\over n+u}={\pi u\over\sin(\pi u)}.
\label{A-u}
\end{equation}
The crossover from (\ref{W-mean1}) to (\ref{W-mean2}) is not sharp,
but occurs over a temperature range of order
\begin{equation}
\delta T\sim T_\ast[\sigma/\ln(R/a)]^{1/2}.
\label{delta-T}
\end{equation}
[The apparent divergence of $A(T/T_\ast)$ at $T=T_\ast$ can be
removed by separating out the contribution from $I_1(1)$.
This procedure yields a full description of the crossover, which 
we shall not elaborate here.]
Note also that, since $A(0)=1$, $\langle W\rangle$ has a finite 
value at $T=0$,
proportional to the density of cells with a negative ground
state pair energy. For small $T$, the excess density
increases as $T^2$.

(ii) Correction to the leading order behavior.
For $T>T_\ast$, the right-hand-side of Eq. (\ref{W-mean2}) appears
as a correction to the leading order behavior, Eq. (\ref{W-mean1}).
The divergence of $A(T/T_\ast)$ at $T=nT_\ast$
signals switching of behavior for $I_1(n)$, and the corresponding
crossover can be analyzed in detail by isolating out the contribution
from this term. 
When the temperature window $2T_\ast<T<T_2$ exists,
there is another correction term to (\ref{W-mean1}) from 
$I_1(2)\simeq y_p^2\langle z^2\rangle\simeq Ny_p^2\exp(2\Delta/T^2)$.
All terms other than these two are shown to be
of order $\langle W\rangle^2$ or smaller.
Since our treatment of the pair-pair interactions is
not accurate enough to produce the coefficient of the $\langle W\rangle^2$
term, these high order corrections will not be considered.

The above analysis shows explicitly that a perturbative calculation
of $\langle W\rangle$ in $y_p$ is dangerous at all temperatures.
Even for $T>T_\ast$, one encounters difficulties when
the calculation is carried out to sufficiently high orders, though
low order terms are well behaved. Such behavior is typical
for a function which has an essential singularity at $y_p=0$.

The calculation of $\langle W^2\rangle$ can be reduced to the
calculation of $\langle W\rangle$ with the help of the identity
\begin{equation}
\langle W^2\rangle=-y_p^2{\partial (y_p^{-1} \langle W\rangle)
\over\partial y_p}.
\label{partial-W2}
\end{equation}
The right-hand-side of the above equation can be evaluated 
using Eqs. (\ref{W-mean1}) and (\ref{W-mean2}) in respective regimes.
For $T<T_\ast$, the leading order result is given by 
(\ref{p0-1}). 
For $T>T_\ast$, the leading order expression of $\langle W\rangle$
is proportional to $y_p$ and hence does not contribute to
$\langle W^2\rangle$.
Going back to Eq. (\ref{W-expand}), and keeping the lowest nonvanishing
terms, we obtain,
\begin{equation}
\langle W^2\rangle\simeq
I_1(2)+
B\Bigl({T\over T_\ast}\Bigr)
{NT_\ast\over(2\pi\Delta)^{1/2}}\exp\Bigl(-{\Delta\over 2T_\ast^2}\Bigr),
\label{p0-2}
\end{equation}
where $B(u)=(1-u)A(u)+u/(u-2)$.
It is seen that, for $T>T_\ast$, $\langle W^2\rangle$ is smaller 
than $\langle W\rangle$
by a factor which decreases as a power-law of $R$.
Equation (\ref{p0-2}) also accounts for the crossover regime
around $T=T_\ast$, and reduces to (\ref{p0-1}) for $T<T_\ast$.

\section{Vortex free energy in the ordered phase}

Following the same idea as in the calculation of $\langle W\rangle$
in Appendix~C,
we may rewrite (\ref{F-v}) as,
\begin{equation}
dF_{\rm v}\simeq
-{T\over R^2}\Bigl(I_3+
\sum_{n=1}^\infty {(-1)^{n-1}\over n}\bigl[I_1(n)+I_2(n)\bigr]\Bigr),
\label{df-expand}
\end{equation}
where 
\begin{equation}
I_3=\int_{1/y_p}^\infty (\ln y_p+\ln z) P(z)dz.
\label{I-3}
\end{equation}

The discussion of Appendix~C indicates that a possible source of
singularity comes from the terms $I_1(n)$ at $T=T_n$,
where the argument of the error function in Eq. (\ref{I-1c})
undergoes relatively rapid change.
A true singularity, however, appears only in the limit
$R\rightarrow\infty$. Since the statistical weight of large
pairs vanish rapidly with $R$, the rapid change of $I_1(n)$ at
large $R$ may not produce a singularity in $F_{\rm v}$.
The following calculation supports this idea.

In terms of $l=\ln(R/a)$, Eq. (\ref{I-1c}) can be expressed as,
\begin{equation}
I_1(n)=\pi y_p^n\exp(-\theta_n l)\bigl[1-{\rm erf}(\delta_n l^{1/2})\bigr]dl
\label{I-1d}
\end{equation}
where
\begin{eqnarray}
\label{theta-n}
\theta_n&=&2\pi nK-4-2\pi\sigma n^2K^2,\\
\nonumber
\\
\label{delta-n}
\delta_n&=&2nT_g/T-2T_g/T_\ast.
\end{eqnarray}
Substituting (\ref{I-1d}) into (\ref{df-expand}),
and integrating over $l$, we obtain the contribution to $F_{\rm v}$
from $I_1(n)$,
\begin{equation}
F_{\rm v, 1}(n)\simeq
(-1)^ny_p^n\Bigl({\pi T\over na^2}\Bigr){[\pi/ (2\sigma)-2]^{-1/2}\over
\delta_n+[\pi/ (2\sigma)-2]^{1/2}}.
\label{F-v-1n}
\end{equation}
In deriving the above equation, we have neglected a weak dependence of
the parameters $\sigma$ and $K=J/T$ on $R$, which is
justified asymptotically in the ordered phase.
It is clear from (\ref{F-v-1n}) that the free energy has no
singularity at $\delta_n=0$ or any other
point in the ordered phase for any $n$.

\begin{figure}
\epsfxsize=8truecm
\epsfbox{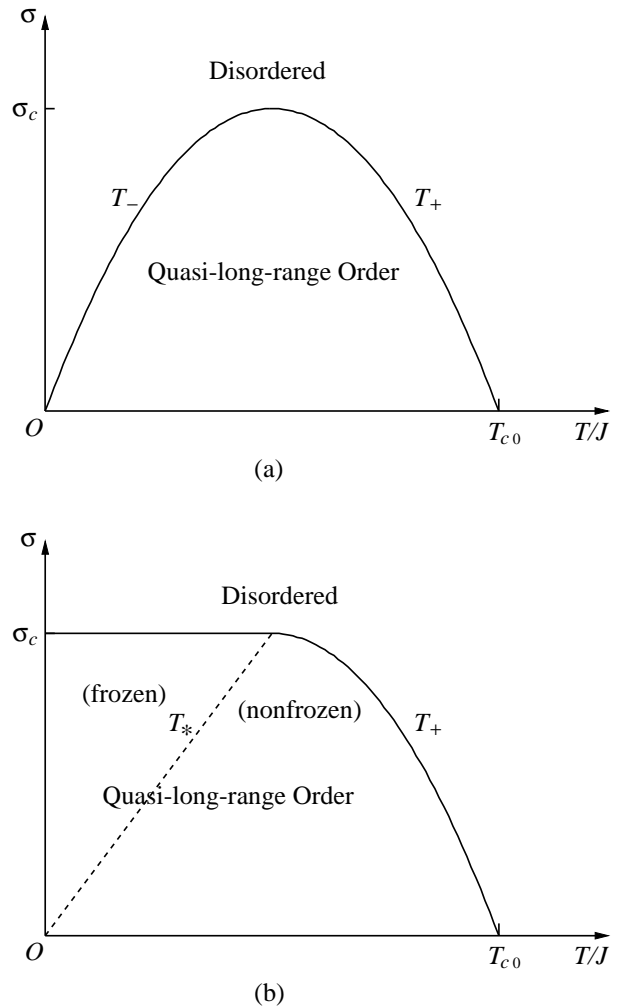}
\smallskip
\narrowtext
\caption{Previously proposed phase diagrams of the
disordered XY model. (a) Order-disorder transition at $T=T_+$
and then again at a reentrance temperature $T=T_-$.
(b) No reentrance transition, but freezing of vortex-pair excitations
below $T_\ast$ (dashed line). 
}
\label{fig1}
\end{figure}

\begin{figure}
\epsfxsize=\linewidth
\epsfbox{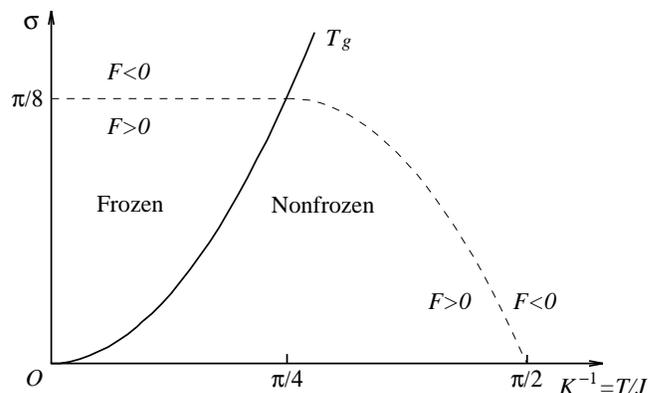}
\narrowtext
\caption{
Phase diagram of a single vortex. A true glass transition
takes place at $T_g$ (solid line). The free energy of the vortex vanishes
along the dashed line.
}
\label{fig2}
\end{figure}

\begin{figure}
\epsfxsize=7.5cm
\epsfbox{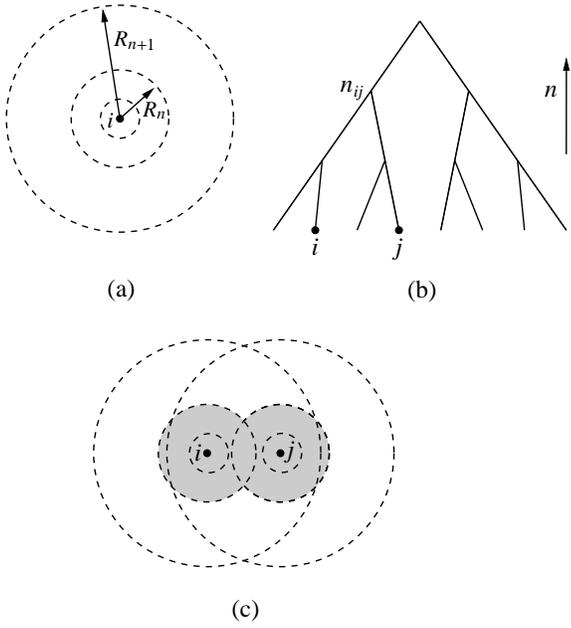}
\smallskip
\narrowtext
\caption{
Cayley tree representation of correlations in the disorder potential
on a single vortex.
(a) Division of the disorder potential into subsums over rings
centered at vortex position $i$. (b) Representation of the potential by
the energy of a path on the Cayley tree. The energy of a path is
the sum over the energies assigned to the nodes.
(c) Two sites $i$ and $j$ pick up nearly identical contribution from
distant quenched disorder, but completely different contribution
from inner shells (shadowed area) surrounding each site.
}
\label{fig3}
\end{figure}

\begin{figure}
\epsfxsize=\linewidth
\epsfbox{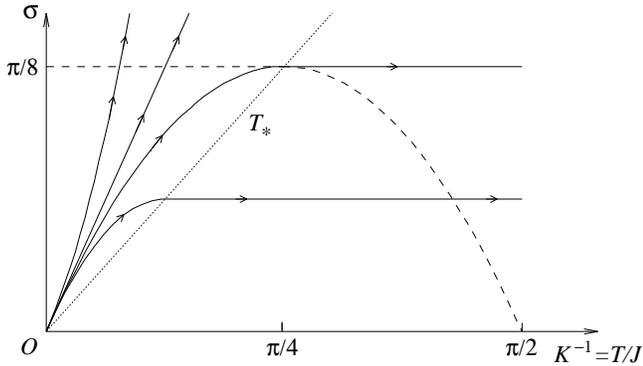}
\smallskip
\narrowtext
\caption{
Renormalization group flows on the $\sigma$-$K^{-1}$ plane.
Solid lines are trajectories of the RG flow, with arrows indicating
flow direction. The flow follows a parabola up to
$T_\ast$ (dotted line), and then joined smoothly by a horizontal line
at $T>T_\ast$.
The dashed line is the line of hyperbolic fixed points
of the RG flow, which gives the phase boundary at $Y=0$.
}
\label{fig4}
\end{figure}

\begin{figure}
\epsfxsize=\linewidth
\epsfbox{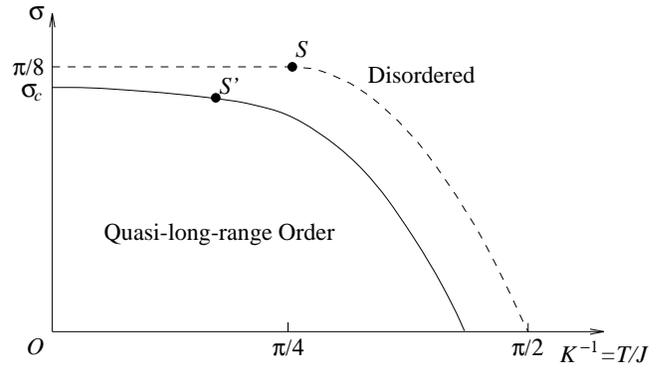}
\medskip
\narrowtext
\caption{
Schematic phase diagram from present work.
Solid line indicates the order-disorder phase boundary
in terms of bare parameters of the model.
It lies inside the region enclosed by the dashed line, which
is the line of phase transition when renormalized values
are plotted. At the special point $S$ and its counterpart
$S'$, where the $T_\ast$-line meets the phase boundary,
a slightly different critical behavior is expected.
See text.
}
\label{fig5}
\end{figure}

\begin{figure}
\epsfxsize=7truecm
\epsfbox{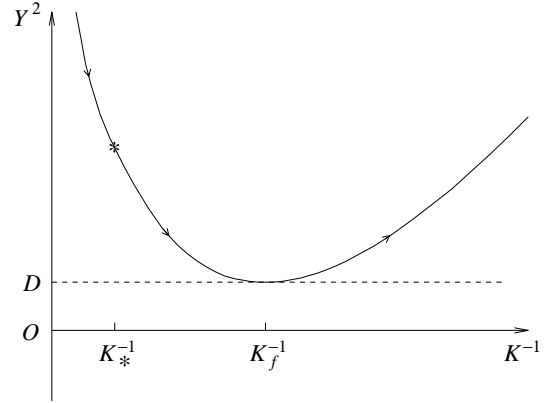}
\medskip
\narrowtext
\caption{
The RG flow trajectory on the $Y^2$-$K^{-1}$ plane.
(This example corresponds to the lowest of the four flow
lines illustrated in Fig. 4.)
At the transition, the bottom of the curve touches
the horizontal axis ($D=0$).
Generically, the curve is quadratic around the minimum at
$K^{-1}=K^{-1}_f$. However, for $K^{-1}_f=K^{-1}_\ast$,
a cubic singularity is found instead.
}
\label{fig6}
\end{figure}
\vskip 5cm
\

\end{multicols}

\begin{references}
\bibitem{note1} Random displacement disorder has the following 
defining property.
When we go around a closed path connecting a set of superconducting grains,
the fluctuation in the magnetic flux enclosed has a variance
proportional to the length of the path, rather than the area
enclosed by the path. In other words, the fluctuation is determined solely
by the displacement of the grains on the path.
This is precisely the situation described by uncorrelated $A_{ij}$'s.
A different situation arises when there is variation in
grain size, in which case
the residual Meissner effect leads to correlated $A_{ij}$'s and a
stronger fluctuation in the flux enclosed.
\bibitem{note3} Equation (\ref{delta-J}) is slightly different from what 
one gets with a straightforward substitution, but is actually more correct
when screening by fellow pairs is taken into account.
\end{references}
\end{document}